# Dynamic Decentralized Algorithms for Cognitive Radio Relay Networks

Siavash Bayat, Raymond H. Y. Louie, Branka Vucetic, Yonghui Li

Centre of Excellence in Telecommunications, School of Electrical and Information Engineering, University of Sydney, Australia

**Abstract**

We propose a distributed spectrum access algorithm for cognitive radio relay networks with multiple primary users (PU) and multiple secondary users (SU). The key idea behind the proposed algorithm is that the PUs negotiate with the SUs on both the amount of monetary compensation, and the amount of time the SUs are either (i) allowed spectrum access, or (ii) cooperatively relaying the PU's data, such that both the PUs' and the SUs' minimum rate requirement are satisfied. The proposed algorithm is shown to be flexible in prioritizing either the primary or the secondary users. We prove that the proposed algorithm will result in the best possible stable matching and is weak Pareto optimal. Numerical analysis also reveal that the distributed algorithm can achieve a performance comparable to an optimal centralized solution, but with significantly less overhead and complexity.

## I. Introduction

Cognitive radio has been proposed as a promising technology to improve the spectral efficiency of wireless networks. This is achieved by allowing unlicensed secondary users (SU) to coexist with licensed primary users (PU) in the same spectrum. This coexistence is facilitated by spectrum access techniques, such as those involving an agreement between the PUs and SUs on an acceptable spectrum access strategy. The key idea is that the PUs are motivated to lease spectrum bands to the SUs in exchange for some form of compensation.

Monetary compensation has been well studied (see e.g., [1–8]), with the predominant approach for spectrum access and performance analysis involving the use of tools from game theory. For example, [7–9] considered a non-cooperative game between the PUs and SUs, while [4, 6] considered a two-stage leader-follower game. For these monetary payment schemes, the PUs are assumed to have sufficient spectrum for leasing to the SUs, such that their own performance requirements are not affected. Authors in [10] also have considered bilateral bargaining among PUs and SUs. In practice, however, the PUs may desire higher data rates than what its current spectrum can provide.

To allow for higher data rates, the use of cooperative relaying has emerged as a powerful technique due to its ability to exploit user diversity and provide high reliability and capacity in wireless networks [11]. This is achieved by the use of intermediate relay nodes to aid transmission between the source and destination nodes. The use of cooperative relaying is particularly advantageous when the direct link between the source and destination is weak, due to, for example, high shadowing.

In this paper, we consider a model where the SUs act as cooperative relays to assist the PUs' transmission in exchange for both spectrum access and monetary compensation, and thus the SUs are effectively providing both a *monetary* and *performance compensation* to the PUs. When only performance compensation is considered, this is commonly referred to as the *overlay* model, and various schemes have been proposed [12–16]. However, these schemes considered the scenario where the increase in PU's performance does not necessarily translate into a satisfactory performance for the SUs, and in some cases, the SUs have limited spectrum access opportunities if the PUs have regular data to transmit [12, 17, 18]. This issue was addressed in [19, 20], where a scheme was proposed which increased the PU's performance while simultaneously satisfying the SU's requirements. However, these papers [12–14, 19–21] considered a simplified scenario with only *one* PU, and did not consider monetary compensation. Besides, some works also considered a simplified scenario with multiple PUs and only one SU.

In this paper, we propose a spectrum access strategy for a more general cognitive radio relay network with *multiple competitive* PUs and *multiple competitive* SUs under the overlay model, which guarantees a *minimum rate requirement* for all matched PUs and SUs. We define a PU and SU as *matched* if the SU cooperatively relays the PU's data, in exchange for spectrum access and monetary compensation.

The majority of current game theory techniques in cognitive radio, which involve multiple PUs and multiple SUs focus on a framework where only users of the same type are the primary decision makers, i.e., either the PUs or the SUs are involved in the game to determine access to the spectrum resources [22]. However, in this approach, the decision makers do not take into account the performance metric of the non-decision users, and thus will lead to unacceptably low performance for these users. To obtain better performance, it is thus desirable for the spectrum access strategy to get all users involved in the decision making process, and thus the interactions between the PUs and SUs should be taken into account, along with the varying performance requirements of these users.

Moreover, what is not actually captured by the normal and basic game and auction models is the fact that different spectrum resources and relaying services that are offered by PUs and SUs causes different performance levels for the SUs and PUs respectively due to random phenomena such as: fading, noise, and shadowing. More specifically, each PU and each SU supplies a spectrum and relaying service with a specific and unique rate performance for a SU and a PU respectively. Thus to model such a complex interaction between multiple spectrum suppliers and multiple relaying service suppliers, a more flexible tool is required. Current algorithms in cognitive radio literature addressing this issue have thus far focused on centralized-approaches, e.g., the double-auction approach in [23]. This approach is clearly not desirable in practice due to the significant amount of overhead required in centralized coordination.

In this paper, we develop a new distributed spectrum access framework based on *auction* and *matching theory*. The proposed distributed algorithms will match the PUs with the SUs based on a preference list, which is unique to each user, and based on *both* monetary and performance compensations. Matching users belonging to two different groups (e.g., a PU group and a SU group) based on preference lists has its roots in the dynamic matching theory [24, 25], which was used to match users based on a preference list, and recently applied to wireless resource allocation problems [26, 27]. However, the algorithms based on the classic matching theory did not consider the possibility of dynamic negotiation between the PUs and the SUs with time-varying requirements, which is desirable for users in next generation wireless access networks.

To address this issue, we propose a novel dynamic negotiation algorithm based on multi-item auction theory [25, 28] and matching theory that requires an intelligent negotiation mechanism, where the PUs and the SUs choose between either prioritizing monetary or performance compensation. Our algorithm determines the matched pairings between PUs and SUs, such that the SU will provide monetary compensation, and relay its paired PU's data in exchange for spectrum access. The key idea behind the algorithm is that the PUs negotiate with the SUs on the amount of monetary compensation, in addition to the time the SUs are either (i) allowed access to the spectrum, or (ii) cooperatively relaying the PU's data, such that both the PUs' and the SUs' minimum rate requirement are satisfied.

We then analyze the performance of the proposed algorithm, showing that it results in the best possible *stable matching*, and is *weak Pareto optimal*. We introduce a utility function, which incorporates both the rate and monetary factors. We demonstrate through numerical analysis that the algorithm can achieve utilities (i) comparable to the utilities achieved by an optimal centralized algorithm, and (ii) significantly greater than the utilities achieved by a random matching algorithm that is mixed with a basic negotiation, while also being able to accomplish a high number of matchings with low overhead and complexity.

When only the rate is important, and monetary compensation is not a priority, we show that our algorithm is flexible in terms of prioritizing either the PUs or SUs, by a simple manipulation of global parameter values. This is in contrast to [29], in which this design flexibility was not achievable. Note that considering this extra monetary parameter is a significant extension to the algorithm presented in [29], as considerable changes to the algorithm are required. Moreover, in contrast to [29], we also present additional analyses including (i) proving that our algorithm results in the best stable matching over all possible stable matchings, (ii) providing an explicit upper bound expression for the number of iterations required for convergence of the proposed algorithm, (iii) presenting an explicit expression for the overhead and (iv) analyzing the complexity performance of our proposed scheme compared to a baseline centralized scheme. Finally, we show that the PUs, which utilize the SUs for cooperative relaying achieve a rate greater than what it would achieve without cooperative relaying, i.e, direct transmission, and thus motivates their participation in the proposed algorithm.

This paper is organized as follows. In Section II, we first describe our system model. We then formulate the optimization problem we are trying to solve in Section III, and present a distributed solution to this problem in Section IV. Finally, we analyze the performance and the implementation aspects of our proposed algorithm in Section V. For convenience, Table I provides a description of some of the parameter values we will be utilizing in this paper.

## II. System Model

We consider an overlay cognitive radio wireless network, comprising of $L_{\rm PU}$ PU transmitter $\{{\rm PT}_i\}_{i=1}^{L_{\rm PU}}$–PU receiver $\{{\rm PR}_i\}_{i=1}^{L_{\rm PU}}$ pairs, with the $\ell$th pair having a rate requirement of $R_{{\rm PU}_\ell,{\rm req}}$, and with each pair occupying a unique spectrum band of constant size. In the same network, there are $L_{\rm SU}$ SU transmitter $\{{\rm ST}_i\}_{i=1}^{L_{\rm SU}}$–SU receiver $\{{\rm SR}_i\}_{i=1}^{L_{\rm SU}}$ pairs, with the $q$th pair having a rate requirement of $R_{{\rm SU}_q,{\rm req}}$, and seeking to obtain access to one spectrum band occupied by a $({\rm PT},{\rm PR})$ pair. We assume that there are $T$ time-slots per transmission frame, and each $(\mathrm{ST}, \mathrm{SR})$ pair has access to a monetary value $C$.

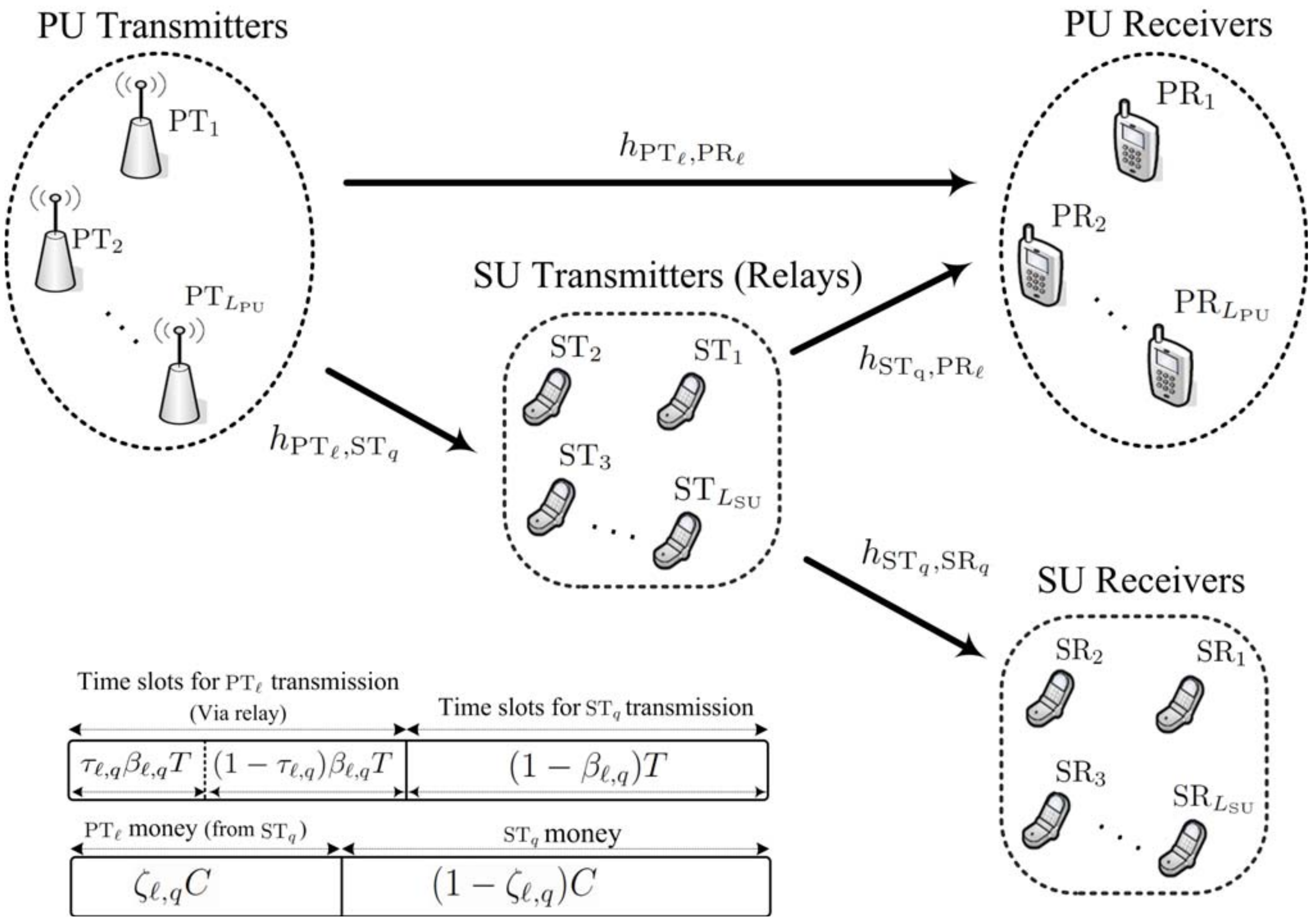

Fig. 1. Secondary user and primary user spectrum-access model. The channel and price and time-slot allocation numbers are indicated for $(\mathrm{PT}_\ell, \mathrm{PR}_\ell)$ and $(\mathrm{ST}_q, \mathrm{SR}_q)$.

Each $\mathrm{PT}$ attempts to grant spectrum access to a unique $(\mathrm{ST}, \mathrm{SR})$ pair, as determined by the various matching algorithms, described in Sections III and IV, in exchange for (i) the $\mathrm{ST}$ cooperatively relaying the $\mathrm{PT}$'s data to the corresponding $\mathrm{PR}$, and (ii) monetary compensation. In particular, without loss of generality (w.l.o.g), let us consider $(\mathrm{PT}_\ell, \mathrm{PR}_\ell)$, whose transmission is relayed by $\mathrm{ST}_q$ during a fraction $\beta_{\ell,q}$ $(0\leq\beta_{\ell,q}\leq1)$ of $T$, whilst also receiving a fraction $\xi_{\ell,q}$ $(0\leq\xi_{\ell,q}\leq1)$ of $C$ from $\mathrm{ST}_q$, as depicted in Figure 1. We will refer to $\xi_{\ell,q}$ and $\beta_{\ell,q}$ as the price and time-slot allocation numbers, respectively, whose exact values will be determined by the matching algorithms described in Sections III and IV.

During the cooperative relaying stage in the initial $\beta_{\ell,q}T$ time-slots, a fraction $\tau_{\ell,q}$ $(0<\tau_{\ell,q}<1)$ is first allocated for $\mathrm{PT}_\ell$ to broadcast its signal to $\mathrm{ST}_q$ and $\mathrm{PR}_\ell$, thus occurring in the first $\beta_{\ell,q}\tau_{\ell,q}T$ time-slots. In the subsequent $\beta_{\ell,q}(1-\tau_{\ell,q})T$ time-slots, $\mathrm{ST}_q$ cooperatively relays the signal from $\mathrm{PT}_\ell$ to $\mathrm{PR}_\ell$. $\mathrm{PR}_\ell$ and then applies maximum ratio combining (MRC) to the signal received from $\mathrm{PT}_\ell$ in the first $\beta_{\ell,q}\tau_{\ell,q}T$time-slots, and the signal received from $\mathrm{ST}_q$ in the subsequent $\beta_{\ell,q}(1-\tau_{\ell,q})T$ time-slots. After this cooperative relaying stage, $\mathrm{PT}_\ell$ ceases transmission, allowing $\mathrm{ST}_q$ to transmit to $\mathrm{SR}_q$ over the spectrum occupied by $(\mathrm{PT}_\ell, \mathrm{PR}_\ell)$ in the final $(1-\beta_{\ell,q})T$ time-slots.

In this paper, we consider the amplify-and-forward (AF) relaying protocol, due to its simple and practical operation, and thus set $\tau_{\ell,q}{=}\frac{1}{2}$. We note, however, that the proposed algorithm is applicable to any relaying protocol, such as the decode-and-forward or compress-and-forward protocol. The AF gain at $\mathrm{ST}_q$ is chosen such that its instantaneous transmission power is constrained to $P_{\mathrm{SU}_q}$.

*A. Utility Functions*

To evaluate the performance of each $(\mathrm{PT}, \mathrm{PR})$ and $(\mathrm{ST}, \mathrm{SR})$ pair, we consider the utility function, which comprises of both rate and monetary factors. The utility function is a concept, which shows the level of satisfaction of a user by combining all parameters to a single number. It represents the net losses and gains of a user [30]. These parameters can be of different types, and by using suitable weights can be combined together.

Utility functions have been widely used in wireless literature to solve various radio resource management problems [31–33]. These papers include combining different parameters, such as price, delay, signal-to-interference and noise ratio, power, rate, error performance, into a single number. Specifically in cognitive radio, the utility function have been used in [1, 4, 9, 34–36].

For $(\mathrm{PT}_\ell, \mathrm{PR}_\ell)$, the achievable instantaneous rate is given by [11]

$$R_{\mathrm{PU}_{\ell,q}}(\beta_{\ell,q}){=}\frac{\beta_{\ell,q}T}{2}\log_2\left(1+\Gamma^{\mathrm{Dir}}_{\mathrm{PR}_\ell}+\Gamma^{\mathrm{Relay}}_{\mathrm{PR}_{\ell,q}}\right) \tag{1}$$

where

$$\Gamma^{\mathrm{Dir}}_{\mathrm{PR}_\ell}{=}\frac{\gamma_{\mathrm{PT}_\ell}||h_{\mathrm{PT}_\ell,\mathrm{PR}_\ell}||^2}{d^{\alpha}_{\mathrm{PT}_\ell,\mathrm{PR}_\ell}} \tag{2}$$

is the receive signal-to-noise-ratio (SNR) at $\mathrm{PR}_\ell$ of the direct signal from $\mathrm{PT}_\ell$,

$$\Gamma^{\mathrm{Relay}}_{\mathrm{PR}_{\ell,q}}{=}\frac{\Gamma_{\mathrm{PT}_\ell-\mathrm{ST}_q}\Gamma_{\mathrm{ST}_q-\mathrm{PR}_\ell}}{\Gamma_{\mathrm{PT}_\ell-\mathrm{ST}_q}\Gamma_{\mathrm{ST}_q-\mathrm{PR}_\ell}+1} \tag{3}$$

is the equivalent receive SNR of the relayed signal at $\mathrm{PR}_\ell$ from $\mathrm{ST}_q$, $\Gamma_{\mathrm{PT}_\ell-\mathrm{ST}_q}{=}\frac{\gamma_{\mathrm{PT}_\ell}||h_{\mathrm{PT}_\ell,\mathrm{ST}_q}||^2}{d^{\alpha}_{\mathrm{PT}_\ell,\mathrm{ST}_q}}$, $\Gamma_{\mathrm{ST}_q-\mathrm{PR}_\ell}{=}\frac{\gamma_{\mathrm{ST}_q}||h_{\mathrm{ST}_q,\mathrm{PR}_\ell}||^2}{d^{\alpha}_{\mathrm{ST}_q,\mathrm{PR}_\ell}}$, $\gamma_{\mathrm{PT}_\ell}{=}\frac{P_{\mathrm{PT}_\ell}}{\sigma^2}$ is the transmit SNR for $\mathrm{PT}_\ell$, $\gamma_{\mathrm{ST}_q}{=}\frac{P_{\mathrm{ST}_q}}{\sigma^2}$ is the transmit SNR for $\mathrm{ST}_q$, $\sigma^2$ is the noise variance, $P_{\mathrm{PT}_\ell}$ is the transmission power at $\mathrm{PT}_\ell$, $h_{\mathrm{PT}_\ell,\mathrm{PR}_\ell}$ and $d_{\mathrm{PT}_\ell,\mathrm{PR}_\ell}$ denote respectively the channel and distance from $\mathrm{PT}_\ell$ to $\mathrm{PR}_\ell$, $h_{\mathrm{PT}_\ell,\mathrm{ST}_q}$ and $d_{\mathrm{PT}_\ell,\mathrm{ST}_q}$ are respectively the channel and distance from $\mathrm{PT}_\ell$ to $\mathrm{ST}_q$, $h_{\mathrm{ST}_q,\mathrm{PR}_\ell}$ and $d_{\mathrm{ST}_q,\mathrm{PR}_\ell}$ are respectively the channel and distance from $\mathrm{ST}_q$ to $\mathrm{PR}_\ell$ and $\alpha$ is the path loss exponent. We also consider a slowly varying block fading channel model with a sufficiently long coherence time that is constant over the transmission frame $T$.

Similar to the utility functions defined in [9, 12, 34–36] we define

$$U_{\mathrm{PU}_{\ell,q}}(\beta_{\ell,q},\xi_{\ell,q}){=}R_{\mathrm{PU}_{\ell,q}}(\beta_{\ell,q})+\bar{c}\xi_{\ell,q}C \tag{4}$$

where $\bar{c}{\in}\mathbb{R}^+$, is a variable with unit defined as: rate per unit monetary value.

For $(\mathrm{ST}_q,\mathrm{SR}_q)$, the achievable instantaneous rate is given by

$$R_{\mathrm{SU}_{q,\ell}}(\beta_{\ell,q}){=}(1-\beta_{\ell,q})T\log_2\left(1+\Gamma_{\mathrm{SR}_{q,\ell}}\right) \tag{5}$$

where

$$\Gamma_{\mathrm{SR}_{q,\ell}}{=}\frac{\gamma_{\mathrm{ST}_q}||h_{\mathrm{ST}_q,\mathrm{SR}_q,\ell}||^2}{d^{\alpha}_{\mathrm{ST}_q,\mathrm{SR}_q}} \tag{6}$$

is the receive SNR at $\mathrm{SR}_q$ of the direct transmission signal from $\mathrm{ST}_q$, $h_{\mathrm{ST}_q,\mathrm{SR}_q,\ell}$ is the channel coefficient from $\mathrm{ST}_q$ to $\mathrm{SR}_q$ in the spectrum band occupied by $(\mathrm{PT}_\ell,\mathrm{PR}_\ell)$ and $d_{\mathrm{ST}_\ell,\mathrm{SR}_q}$ is the distance from $\mathrm{ST}_q$ to $\mathrm{SR}_q$. The utility for $(\mathrm{ST}_q,\mathrm{SR}_q)$ is thus given by

$$U_{\mathrm{SU}_{q,\ell}}(\beta_{\ell,q},\xi_{\ell,q}){=}R_{\mathrm{SU}_{q,\ell}}(\beta_{\ell,q})-\bar{k}\xi_{\ell,q}C, \tag{7}$$

where $\bar{k}{\in}\mathbb{R}^+$ is a variable with unit defined as: rate per unit monetary value. We consider a general case where $\bar{c}$ and $\bar{k}$ have the same units, but their values may be different. This is to allow for flexibility in the algorithm design, as will be discussed in Section V.

## III. Problem Formulation

In this section, we describe the optimization problem we aim to address. To proceed, we introduce some notations. We first define the primary and secondary user sets, respectively, as $\mathcal{P}{=}\{\mathrm{PU}_\ell{=}(\mathrm{PT}_\ell,\mathrm{PR}_\ell)\}_{\ell=1}^{L_{\mathrm{PU}}}$ and $\mathcal{S}{=}\{\mathrm{SU}_q{=}(\mathrm{ST}_q,\mathrm{SR}_q)\}_{q=1}^{L_{\mathrm{SU}}}$. Moreover, we define a $L_{\mathrm{PU}}\times L_{\mathrm{SU}}$ matching matrix $\mathbf{M}$, with $m_{i,j}{=}1$ if $\mathrm{PU}_i$ is matched with $\mathrm{SU}_j$, and $m_{i,j}{=}0$ otherwise, where the notation $m_{i,j}$ denotes the $(i,j)$th entry of matrix $\mathbf{M}$. From this matrix, we introduce an injective function $\mu{:}(\mathcal{P}\cup\mathcal{S}){\to}(\mathcal{P}\cup\mathcal{S}\cup\{\varnothing\})$, such that

(a) $\mu(\mathrm{PU}_\ell){\in}(\mathcal{S}\cup\{\varnothing\})$,

TABLE I
PARAMETER DESCRIPTIONS

| Notation | Description |
|---|---|
| $\xi_{\ell,q}$ | Price allocation number |
| $\beta_{\ell,q}$ | Time-slot allocation number |
| $T$ | Length of transmission frame |
| $C$ | Total money each SU has access to in each transmission frame |
| $\tau_{\ell,q}$ | Time-slot fraction that $\mathrm{PT}_{\ell}$ transmits its signal to $\mathrm{ST}_{q}$ |
| $\alpha$ | Path loss exponent |
| $\Gamma_{\mathrm{PR}_{\ell}}^{\mathrm{Dir}}$ | Receive SNR at $\mathrm{PR}_{\ell}$ of the direct transmission signal from $\mathrm{PT}_{\ell}$ |
| $\Gamma_{\mathrm{PR}_{\ell,q}}^{\mathrm{Relay}}$ | Equivalent receive SNR of the relayed signal at $\mathrm{PR}_{\ell}$ from $\mathrm{ST}_{q}$ |
| $\Gamma_{\mathrm{SR}_{q,\ell}}$ | Receive SNR at $\mathrm{SR}_{q}$ of the direct transmission signal from $\mathrm{ST}_{q}$ |
| $\gamma_{\mathrm{PT}_{\ell}}$ | Transmit signal-to-noise-ratio for $\mathrm{PT}_{\ell}$ |
| $\gamma_{\mathrm{ST}_{q}}$ | Transmit signal-to-noise-ratio for $\mathrm{ST}_{q}$ |
| $\bar{c}$ | Rate per unit monetary value for PT |
| $k$ | Rate per unit monetary value for ST |
| $U_{\mathrm{PU}_{\ell,q}}$ | $\mathrm{PU}_{\ell}$'s utility while cooperating with $\mathrm{ST}_{q}$ |
| $U_{\mathrm{SU}_{q,\ell}}$ | $\mathrm{SU}_{q}$'s utility while cooperating with $\mathrm{PU}_{\ell}$ |
| $\mathbf{M}$ | Matching matrix |
| $\mathbf{B}$ | Time-slot allocation matrix |
| $\mathbf{G}$ | Price allocation matrix |
| $\mathrm{PULIST}_{\ell}$ | $\mathrm{PT}_{\ell}$'s preference list |
| $\mathrm{SULIST}_{q}$ | $\mathrm{ST}_{q}$'s preference list |
| $\delta$ | PT's price step number |
| $\epsilon$ | PT's time-slot step number |

(b) $\mu(\mathrm{SU}_q) \in (\mathcal{P} \cup \{\varnothing\})$,

(c) $\mu(\mathrm{SU}_q)=\mathrm{PU}_\ell$ and $\mu(\mathrm{PU}_\ell)=\mathrm{SU}_q$ if $m_{\ell,q}=1$, for $\ell=1,\ldots,L_{\mathrm{PU}}$ and $q=1,\ldots,L_{\mathrm{SU}}$,

(d) $\mu(\mathrm{SU}_q)=\varnothing$ if $m_{\ell,q}=0$, for $\ell=1,\ldots,L_{\mathrm{PU}}$,

(e) $\mu(\mathrm{PU}_\ell)=\varnothing$ if $m_{\ell,q}=0$, for $q=1,\ldots,L_{\mathrm{SU}}$.

In the above definition, (a) implies that a PU is matched to a single SU or no PU, i.e., $\mu(\mathrm{PU}_\ell)=\varnothing$, (b) implies that a SU is matched to a single PU or no source, i.e., $\mu(\mathrm{SU}_q)=\varnothing$, (c) implies that if $\mathrm{PU}_\ell$ is matched to $\mathrm{SU}_q$, then $\mathrm{SU}_q$ is also matched to $\mathrm{PU}_\ell$ and $m_{\ell,q}=1$, (d) implies that if $m_{\ell,q}=0$ then $\mathrm{SU}_q$ is not matched and (e) implies that if $m_{\ell,q}=0$ then $\mathrm{PU}_\ell$ is not matched .

We also define an $L_{\mathrm{PU}} \times L_{\mathrm{SU}}$ price allocation matrix $\mathbf{G}$ with $g_{i,j}=\xi_{i,j}$, and an $L_{\mathrm{PU}} \times L_{\mathrm{SU}}$ time-slot allocation matrix $\mathbf{B}$ with $b_{i,j}=\beta_{i,j}$, and where $g_{i,j}=b_{i,j}=0$ if $m_{i,j}=0$. We denote the price and time-slot allocation matrices with continuous elements as $\mathbf{G}^{\mathrm{cont}}$ and $\mathbf{B}^{\mathrm{cont}}$ respectively. Mathematically, this implies that the elements of $\mathbf{G}^{\mathrm{cont}}$ and $\mathbf{B}^{\mathrm{cont}}$, respectively take values from the sets $\{g_{i,j}^{\mathrm{cont}}=\xi_{i,j}\in \mathbb{R}:0\leq\xi_{i,j}\leq 1\}$ and $\{b_{i,j}^{\mathrm{cont}}=\beta_{i,j}\in\mathbb{R}:0\leq\beta_{i,j}\leq 1\}$.

Now the main goal for each primary and secondary user is to ensure their minimum rate requirements are satisfied. When this is achieved, the secondary goal is to maximize their utility functions. Note that the secondary goals for the primary and secondary users cannot be achieved simultaneously, as a higher utility for the primary user will result in a lower utility for the matched secondary user, and vice-versa. It is natural, and often considered in literature (see e.g. [34]), to give preference to the primary users, i.e. focus on maximizing the primary users' utility. As such, we now present the optimization problem:

$$\begin{aligned}
&\{\mathbf{M}^{\text{opt}}, \mathbf{B}^{\text{opt}}, \mathbf{G}^{\text{opt}}\} \\
&\quad = \arg \max_{\{\mathbf{M}, \mathbf{B}^{\text{cont}}, \mathbf{G}^{\text{cont}}\}} \sum_{\ell=1}^{L_{\text{PU}}} \sum_{q=1}^{L_{\text{SU}}} m_{\ell,q} U_{\text{PU}_{\ell,q}}(\xi_{\ell,q}, \beta_{\ell,q}) \\
&\text{s.t.:} (a) R_{\text{PU}_{\ell,\mu^\dagger(\ell)}}(\beta_{\ell,\mu^\dagger(\ell)}) \geqslant R_{\text{PU}_{\ell,\text{req}}}, \forall \ell \in \{1,2,...,L_{\text{PU}}\} \\
&\quad (b) R_{\text{SU}_{q,\mu^\dagger(q)}}(\beta_{\mu^\dagger(q),q}) \geqslant R_{\text{SU}_{q,\text{req}}}, \forall q \in \{1,2,...,L_{\text{SU}}\} \\
&\quad (c) R_{\text{SU}_{q,\mu^\dagger(q)}}(\beta_{\mu^\dagger(q),q}) - \xi_{\mu^\dagger(q),q} \bar{k} C \geqslant 0, \\
&\qquad \forall q \in \{1,2,...,L_{\text{SU}}\} \\
&\quad (d) \textstyle\sum_{\ell=1}^{L_{\text{PU}}} m_{\ell,q} \leq 1, \forall q \in \{1,2,...,L_{\text{SU}}\} \\
&\quad (e) \textstyle\sum_{q=1}^{L_{\text{SU}}} m_{\ell,q} \leq 1, \forall \ell \in \{1,2,...,L_{\text{PU}}\} \\
&\quad (f) 0 \leqslant \xi_{\ell,q} \leqslant 1, \forall q \in \{1,2,...,L_{\text{SU}}\}, \forall \ell \in \{1,2,...,L_{\text{PU}}\} \\
&\quad (g) 0 \leqslant \beta_{\ell,q} \leqslant 1, \forall q \in \{1,2,...,L_{\text{SU}}\}, \forall \ell \in \{1,2,...,L_{\text{PU}}\}
\end{aligned} \tag{8}$$

where $\mu^\dagger(p_\ell) = q$ if $\mu(p_\ell) = s_q$, and $\mu^\dagger(s_q) = \ell$ if $\mu(s_q) = p_\ell$.

Conditions $(a)$ and $(b)$ ensure that the minimum required rate for the PUs and SUs are satisfied[1], respectively. Condition $(c)$ ensures that the SUs always receive a positive utility. Conditions $(d)$ and $(e)$ respectively ensure that each PU will only be matched with one SU, and vice-versa. Finally, conditions $(f)$ and $(g)$, respectively, ensure that the price and time-slot allocation values are kept within their bounds.

In practice, a centralized controller is required to solve the optimization problem in (8). However, there are three key issues regarding this approach:

- *Overhead:* The centralized controller will require the feedback on channel conditions and minimum rate requirements from each primary and secondary user. Moreover, after the optimization problem is solved, the resultant matching between PUs and SUs and price and time-slot allocation numbers will then have to be transmitted to the corresponding users. The amount of overhead required for this increases with the number of users, and can be quite high, rendering it impractical.

[1] From the output of the optimization problem, it can be shown that the final time slot allocation numbers are chosen such that the achievable rate for the matched SUs is equivalent to their minimum rate requirement, and the price-time slot allocation numbers are chosen such that the utility for each matched SU is zero. We can thus present an alternate formulation accordingly, however, we leave the formulation as described to make it clear that the SU's rate and utility requirements are satisfied.

- *Complexity:* The optimization problem is non-linear, and requires an exhaustive search over all possible matching, price and time-slot allocation combinations. Such a problem is known to be NP-hard [5].
- *Selfish Users:* We assume all the primary and secondary users are selfish[2], which means their goal is to always maximize their own utilities, then the outcome of the optimization problem may not be in the best interests of at least one of these users. Selfishness is considered as an inherent behavior of distributed and intelligent users. There are also privacy issues for which a centralized approach may not be ideal.

To address these issues, we propose a distributed low-complexity algorithm, which accounts for selfish users. As we will demonstrate in Section V, our algorithm can achieve a performance close to the solution of the optimization problem in (8) for practical system parameters.

## IV. Proposed Distributed Matching Algorithm

In this section, we describe the proposed algorithm, which determines spectrum access for each $(\mathrm{PT}, \mathrm{PR})$ and $(\mathrm{ST}, \mathrm{SR})$ pair.

### A. Received SNR Assumptions

We first describe two scenarios that will be considered in the proposed algorithm, characterized by different assumptions on the received SNR at the transmitters and receivers.

*1) Complete Received SNR:* In the first scenario, $\mathrm{PT}_\ell$ has perfect knowledge of the instantaneous received SNRs in $\Gamma_{\mathrm{PR}_\ell}^{\mathrm{Dir}}$ and $\left\{\Gamma_{\mathrm{PT}_\ell-\mathrm{ST}_q}, \Gamma_{\mathrm{ST}_q-\mathrm{PR}_\ell}\right\}_{q=1}^{L_{\mathrm{SU}}}$. Moreover, $\mathrm{ST}_q$ has perfect knowledge of the instantaneous received SNRs in the expressions $\left\{\Gamma_{\mathrm{SR}_{q,\ell}}\right\}_{\ell=1}^{L_{\mathrm{PU}}}$. As such, $\mathrm{PT}_\ell$ and $\mathrm{ST}_q$ are able to respectively calculate their *instantaneous* rates in (1) and (5).

*2) Partial Received SNR:* In the second scenario, $\mathrm{PT}_\ell$ has knowledge of the average received SNRs in the term $\left\{\frac{\gamma_{\mathrm{ST}_q}}{d_{\mathrm{ST}_q,\mathrm{PR}_\ell}^{\alpha}}\right\}_{q=1}^{L_{\mathrm{SU}}}$ and the instantaneous received SNRs in the terms $\Gamma_{\mathrm{PR}_\ell}^{\mathrm{Dir}}$ and $\left\{\Gamma_{\mathrm{PT}_\ell-\mathrm{ST}_q}\right\}_{q=1}^{L_{\mathrm{SU}}}$. Moreover, $\mathrm{ST}_q$ has perfect knowledge of the instantaneous received SNRs in the term $\left\{\Gamma_{\mathrm{SR}_{q,\ell}}\right\}_{\ell=1}^{L_{\mathrm{PU}}}$. As such, $\mathrm{PT}_\ell$ is able to calculate its instantaneous conditional rate, given by the expectation of the rate in (1), with respect to $\{h_{\mathrm{PT}_\ell,\mathrm{ST}_q}\}_{q=1}^{L_{\mathrm{SU}}}$, while $\mathrm{ST}_q$ is able to calculate its *instantaneous* rate in (5).

[2]Selfish users are a common assumption in cognitive radio literature [30].

For both complete and partial received SNR scenarios, note that each $\mathrm{PT}$ and $\mathrm{ST}$ does *not* have knowledge of the instantaneous received SNRs corresponding respectively to the other $\mathrm{PTs}$ and $\mathrm{STs}$. Moreover, the instantaneous received SNRs can be obtained through standard channel estimation techniques.

*B. Users Preference Lists*

Each $\mathrm{PT}$ has a preference list of $\mathrm{STs}$ which can cooperatively relay the $\mathrm{PT}$'s message such that it obtains a rate greater than its minimum rate requirement. In particular, the preference list for $\mathrm{PT}_\ell$ is given by

$$\mathrm{PULIST}_\ell = \{(\mathrm{ST}_{\phi_\ell(j)}, \mathrm{SR}_{\phi_\ell(j)})\}_{j=1}^{K_\ell} \tag{9}$$

where $\phi_\ell(\cdot)$ is a function $\phi_\ell : \{1, \ldots, K_\ell\} \rightarrow \{1, \ldots, L_{\mathrm{SU}}\}$ satisfying

$$\left\{ R_{\mathrm{PU}_{\ell,\phi_\ell(q)}}(\xi_{\ell,\phi_\ell(q)}, \beta_{\ell,\phi_\ell(q)}) \geqslant R_{\mathrm{PU}_\ell,\mathrm{req}} \right\}_{q=1}^{K_\ell} \tag{10}$$

for the complete received SNR scenario, and the conditions

$$\left\{ \mathrm{E}_{h_{\mathrm{ST}_{\phi_\ell(q)},\mathrm{PR}_{\phi_\ell(q)}}} \left[ R_{\mathrm{PU}_{\ell,\phi_\ell(q)}}(\xi_{\ell,\phi_\ell(q)}, \beta_{\ell,\phi_\ell(q)}) \right] \geqslant R_{\mathrm{PU}_\ell,\mathrm{req}} \right\}_{q=1}^{K_\ell} \tag{11}$$

for the partial received SNR scenario. The function $\phi_\ell(\cdot)$ also satisfies the ordering $U_{\mathrm{PU}_{\ell,\phi_\ell(1)}}(\xi_{\ell,\phi_\ell(1)}, \beta_{\ell,\phi_\ell(1)}) > \ldots > U_{\mathrm{PU}_{\ell,\phi_\ell(K_\ell)}}(\xi_{\ell,\phi_\ell(K_\ell)}, \beta_{\ell,\phi_\ell(K_\ell)})$, implying that the first ST in the list provides the largest utility. Moreover, $K_\ell$ is the number of $(\mathrm{ST}, \mathrm{SR})$ pairs satisfying these conditions.

Similarly, each $\mathrm{ST}$ has a preference list of $\mathrm{PTs}$ which, if it transmits in the spectrum band occupied by the $(\mathrm{PT}, \mathrm{PR})$ pair in the list, obtains a rate greater than its minimum rate requirement and a utility greater or equal to zero. In particular, the preference list for $\mathrm{ST}_q$ is given by

$$\mathrm{SULIST}_q = \left\{ (\mathrm{PT}_{\psi_q(\ell)}, \mathrm{PR}_{\psi_q(\ell)}) \right\}_{\ell=1}^{V_q} \tag{12}$$

where $\psi_q(\cdot)$ is a function $\psi_q : \{1, \ldots, V_q\} \rightarrow \{1, \ldots, L_{\mathrm{PU}}\}$ satisfying the conditions

$$\left\{ R_{\mathrm{SU}_{q,\psi_q(\ell)}}(\beta_{q,\psi_q(\ell)}) \geqslant R_{\mathrm{SU}_q,\mathrm{req}} \right\}_{\ell=1}^{V_q} \tag{13}$$

and

$$\left\{ U_{\mathrm{SU}_{q,\psi_q(\ell)}}(\xi_{q,\psi_q(V_\ell)}, \beta_{q,\psi_q(\ell)}) \geqslant 0 \right\}_{\ell=1}^{V_q} \tag{14}$$

with the ordering $U_{\mathrm{SU}_{q,\psi_q(1)}}(\xi_{q,\psi_q(1)}, \beta_{q,\psi_q(1)}) > \ldots > U_{\mathrm{SU}_{q,\psi_q(V_q)}}(\xi_{q,\psi_q(V_q)}, \beta_{q,\psi_q(V_q)})$. The ordering thus implies that the first PT in the list provides the largest utility. Moreover, $V_\ell$ is the number of $(\mathrm{PT}, \mathrm{PR})$ pairs satisfying these conditions.

In practice, the instantaneous channels can be measured by utilizing common channel estimation techniques [37]. One possibility is for this channel estimation to be sent via control channels, as considered in [1, 12].

### *C. Proposed Algorithm to Determine the Matching, Price and Time-slot Allocation Matrices*

The key idea of the proposed algorithm is that each $(\mathrm{PT}, \mathrm{PR})$ pair trades with the $(\mathrm{ST}, \mathrm{SR})$ pair, which provides the highest utility, through both cooperative relaying and monetary payment. This trading will be done by negotiating on the price and time-slot allocation numbers $\{\xi_{\ell,q}, \beta_{\ell,q}\}_{\ell=1}^{L_{\mathrm{PU}}}{}_{q=1}^{L_{\mathrm{SU}}}$. We say $\mathrm{PT}_\ell$ makes an *offer* of $(\xi_{\ell,q}, \beta_{\ell,q})$ to $\mathrm{ST}_q$ to imply that $\mathrm{PT}_\ell$ is willing to allow $\mathrm{ST}_q$ to transmit, in exchange for $\mathrm{ST}_q$ (i) cooperatively relaying $\mathrm{PT}_\ell$'s message with time slot allocation number $\beta_{\ell,q}$ and, (ii) providing a monetary payment with price allocation number $\xi_{\ell,q}$.

The specific details of the main algorithm are given in Table II. Note that the main algorithm calls upon the function[3] 'Proposal Update Unit (PUU)', denoted as $\mathrm{PUU}(\cdot,\cdot,\cdot)$, and detailed in Table III. Note that the PUU provides an intelligent response for each PU after its offer is rejected by a SU.

To summarize the main algorithm (MA), each PT will first make an offer to the ST, which is first in its preference list (MA-Step 2-1). The ST will then check if the offering PT is in it's preference list (MA-Step 2-2-1). If it is, and the ST is already matched with another PT, the ST has two choices: (a) if the offering PT can provide a better utility than the ST's current matching, then the ST will reject its current matching in favor of the new matching (MA-Step 2-2-1-a-i), or (b) if the offering PT can not provide a better utility than the ST's current matching, the ST will reject the PT's offer (MA-Step 2-2-1-a-ii). If the ST is not matched, then the ST will be matched with the offering PT (MA-Step 2-2-1-b). If the offering PT is not in the ST's preference list, the ST will reject the offering PT (MA-Step 2-2-2). The algorithm will then repeat this procedure with each PT until no more matchings are possible.

Note that if the ST rejects a PT, then the proposal update unit (PUU) will be activated, and the PT will either (i) decrease its price allocation number by a price step number $\delta$ (PUU-3), or (ii)

[3]Note that the PUU is required in the algorithm as we are considering both monetary and rate factors in the utility function. This is in contrast to our conference paper [29], which did not have a PUU as monetary factors were not considered.

TABLE II

Main Algorithm (MA)

**Step 1: Initialization.** (Parameter setting and list construction)

**(1)** Set $(\xi_{k,q}, \beta_{k,q}){=}(\xi_{\text{init}}, \beta_{\text{init}})$, for $k{=}1,\dots,L_{\text{PU}}$, $eq{=}1,\dots,L_{\text{SU}}$.
**(2)** Set the price step number $\delta$.
**(3)** Set the time-slot step number $\epsilon$.
**(4)** Construct PUs's preference list, $\text{PULIST}_k$, based on $(\xi_{\text{init}}, \beta_{\text{init}})$, for $k{=}1,\dots,L_{\text{PU}}$.
**(5)** Construct SUs's preference list, $\text{SULIST}_q$, based on $(\xi_{\text{init}}, \beta_{\text{init}})$, for $q{=}1,\dots,L_{\text{SU}}$.
**(6)** Construct the list of all PTs which are not matched, denoted by $\text{MATCH}{=}\{\text{PT}_1,\dots,\text{PT}_{L_{\text{PU}}}\}$.
**(7)** Set the indexing parameter $\ell{=}1$.

**Step 2: Find a suitable ST for each PT**(PUs and SUs dynamic negotiation)

**(1)** $\text{PT}_\ell$ makes an offer of $(\xi_{\ell,q}, \beta_{\ell,q})$ to the first ST in its preference list $\text{PULIST}_\ell$. Let us denote this ST as $\text{ST}_q$. If $(\xi_{\ell,q}, \beta_{\ell,q}){\neq}(\xi_{\text{init}}, \beta_{\text{init}})$, then $\text{ST}_q$ updates its preference list based on $(\xi_{\ell,q}, \beta_{\ell,q})$ and reorder its preference list based on the new update.

**(2)**

1) If $\text{PT}_\ell$ is in $\text{ST}_q$'s preference list, then
   a) If $\text{ST}_q$ is already matched to a primary transmitter, denoted by $\text{PT}_{\text{curr}}$, then
      i) If $\text{PT}_\ell$ is higher up than $\text{PT}_{\text{curr}}$ on $\text{ST}_q$'s preference list $\text{SULIST}_q$ then
         - $\text{SU}_q$ and $\text{PU}_\ell$ are matched. Remove $\text{PU}_\ell$ from MATCH.

         - Update $(\xi_{\text{curr},q}, \beta_{\text{curr},q}, \text{MATCH}, \text{PULIST}_{\text{curr}})$ by executing the Proposal Update Unit, ie., $(\xi_{\text{curr},q}, \beta_{\text{curr},q}, \text{MATCH}, \text{PULIST}_{\text{curr}})$ $=\text{PUU}\,(\xi_{\text{curr},q}, \beta_{\text{curr},q}, \text{MATCH}, \text{PULIST}_{\text{curr}})$.
      ii) Else
         - Update $(\xi_{\ell,q}, \beta_{\ell,q}, \text{MATCH}, \text{PULIST}_\ell)$ by executing the Proposal Update Unit, ie., $(\xi_{\ell,q}, \beta_{\ell,q}, \text{MATCH}, \text{PULIST}_\ell)$ $=\text{PUU}\,(\xi_{\ell,q}, \beta_{\ell,q}, \text{MATCH}, \text{PULIST}_\ell)$.
   b) Else
      - $\text{SU}_q$ and $\text{PU}_\ell$ are matched. Remove $\text{PU}_\ell$ from MATCH.
2) Else
   - Update $\xi_{\ell,q}, \beta_{\ell,q}, \text{MATCH}, \text{PULIST}_\ell$ by executing the Proposal Update Unit, ie., $(\xi_{\ell,q}, \beta_{\ell,q}, \text{MATCH}, \text{PULIST}_\ell)$ $=\text{PUU}\,(\xi_{\ell,q}, \beta_{\ell,q}, \text{MATCH}, \text{PULIST}_\ell)$.
3) If $\text{PULIST}_\ell{=}\emptyset$, remove $\text{PT}_\ell$ from MATCH, and go to Step 2-3, else go to Step 2-1.

**(3)** If no more matchings are possible, i.e., $\text{MATCH}{=}\emptyset$, then go to Step 4. Else go to Step 2-1 for each PT which is not matched, i.e., go to Step 2-1 with $\text{PT}_\ell$ being the PT corresponding to the first entry in MATCH.

**Step 4: End of the algorithm**

decrease its time slot allocation number by a time slot-step number $\epsilon$ (PUU-4), depending on which option maximizes the PT's utility, and assuming a positive price and time-slot allocation number and the minimum data rate requirement for the PT is satisfied.

We observe that the proposed algorithm produces a matching, price allocation, and time-slot allocation matrix. Note that the price and time-slot allocation matrix have non-zero entries, which

TABLE III

Proposal Update Unit (PUU)

**Input**

**(1)** Current price allocation number $\xi_{\ell,q}^{\mathrm{Old}}$.
**(2)** Current time-slot allocation number $\beta_{\ell,q}^{\mathrm{Old}}$.
**(3)** Current $\mathrm{MATCHLIST}^{\mathrm{Old}}$
**(4)** Current $\mathrm{PU}_{\ell}^{\mathrm{Old}}$'s preference list $\mathrm{PULIST}_{\ell}^{\mathrm{Old}}$.

**Output**

**(1)** Updated price allocation number $\xi_{\ell,q}^{\mathrm{New}}$.
**(2)** Updated time-slot allocation number $\beta_{\ell,q}^{\mathrm{New}}$.
**(3)** Updated $\mathrm{MATCHLIST}^{\mathrm{New}}$
**(4)** Updated $\mathrm{PU}_{\ell}$'s preference list $\mathrm{PULIST}_{\ell}^{\mathrm{New}}$.

**Procedure**

**(1)** If $\xi_{\ell,q}^{\mathrm{Old}}-\delta\leq 0$
1) $\beta_{\ell,q}^{\mathrm{New}}=\max\left(\beta_{\ell,q}^{\mathrm{Old}}-\epsilon,0\right)$.
2) $\xi_{\ell,q}^{\mathrm{New}}=\xi_{\ell,q}^{\mathrm{Old}}$.

**(2)** Else if $\xi_{\ell,q}^{\mathrm{Old}}-\delta>0$ and $R_{\mathrm{PU}_{\ell,q}}\left(\beta_{\ell,q}^{\mathrm{Old}}-\epsilon,\xi_{\ell,q}\right)\leq R_{\mathrm{PU}_{\ell},\mathrm{req}}$
1) $\beta_{\ell,q}^{\mathrm{New}}=\beta_{\ell,q}^{\mathrm{Old}}$.
2) $\xi_{\ell,q}^{\mathrm{New}}=\xi_{\ell,q}^{\mathrm{Old}}-\delta$.

**(3)** Else if $U_{\mathrm{PU}_{\ell,q}}\left(\beta_{\ell,q}^{\mathrm{Old}},\xi_{\ell,q}^{\mathrm{Old}}-\delta\right)<U_{\mathrm{PU}_{\ell,q}}\left(\beta_{\ell,q}^{\mathrm{Old}}-\epsilon,\xi_{\ell,q}^{\mathrm{Old}}\right)$
1) $\beta_{\ell,q}^{\mathrm{New}}=\beta_{\ell,q}^{\mathrm{Old}}-\epsilon$.
2) $\xi_{\ell,q}^{\mathrm{New}}=\xi_{\ell,q}^{\mathrm{Old}}$.

**(4)** Else $\left(U_{\mathrm{PU}_{\ell,q}}\left(\beta_{\ell,q}^{\mathrm{Old}},\xi_{\ell,q}^{\mathrm{Old}}-\delta\right)\right.$
$\left.\geq U_{\mathrm{PU}_{\ell,q}}\left(\beta_{\ell,q}^{\mathrm{Old}}-\epsilon,\xi_{\ell,q}^{\mathrm{Old}}\right)\right)$
1) $\beta_{\ell,q}^{\mathrm{New}}=\beta_{\ell,q}^{\mathrm{Old}}$.
2) $\xi_{\ell,q}^{\mathrm{New}}=\xi_{\ell,q}^{\mathrm{Old}}-\delta$.

**(5)** Update the old preference list $\mathrm{PULIST}_{\ell}^{\mathrm{Old}}$ for $(\mathrm{PT}_{\ell},\mathrm{PR}_{\ell})$ based on the updated $\left(\xi_{\ell,q}^{\mathrm{New}},\beta_{\ell,q}^{\mathrm{New}}\right)$, to form a new preference list $\mathrm{PULIST}_{\ell}^{\mathrm{New}}$.
**(6)** Update the old $\mathrm{MATHLIST}^{\mathrm{Old}}$ by putting $\mathrm{PU}_{\ell}$ at the end of the list, to form a new $\mathrm{MATCHLIST}^{\mathrm{New}}$.

take values from a discrete set, in contrast to the continuous set considered in the optimization problem in (8). This is due to the update procedure, where the price and time-slot allocation numbers change according to the price and time-slot step numbers $\delta$ and $\epsilon$. We denote the price and time-slot allocation matrices with discrete elements corresponding to the particular $\xi_{\mathrm{init}}$, $\beta_{\mathrm{init}}$, $\delta$ and $\epsilon$ values as $\mathbf{G}^{\mathrm{disc}}(\xi_{\mathrm{init}},\delta)$ and $\mathbf{B}^{\mathrm{disc}}(\beta_{\mathrm{init}},\epsilon)$ respectively. Mathematically, the elements of $\mathbf{G}^{\mathrm{disc}}(\xi_{\mathrm{init}},\delta)$ and $\mathbf{B}^{\mathrm{disc}}(\beta_{\mathrm{init}},\epsilon)$ respectively take values from the sets $\{g_{i,j}^{\mathrm{disc}}=\xi_{i,j}\in\xi_{\mathrm{init}}-m\delta{:}m=1,\ldots,\lfloor\frac{\xi_{\mathrm{init}}}{\delta}\rfloor\}$ and $\{b_{i,j}^{\mathrm{disc}}=\beta_{i,j}\in\beta_{\mathrm{init}}-m\epsilon{:}m=1,\ldots,\lfloor\frac{\beta_{\mathrm{init}}}{\epsilon}\rfloor\}$.

## V. Performance and Implementation Analysis

We now analyze the performance of the proposed algorithm, and consider related implementation issues. We first present some assumptions we will be considering in the analysis. To demonstrate that the $(\mathrm{PT},\mathrm{PR})$ pairs are motivated to participate in the proposed algorithm, we set the minimum rate requirement of each $(\mathrm{PT},\mathrm{PR})$ pair to be the rate of the direct $\mathrm{PT}$ to $\mathrm{PR}$ link. This is given for $(\mathrm{PT}_{\ell},\mathrm{PR}_{\ell})$ by $R_{\mathrm{PT}_{\ell},\mathrm{PR}_{\ell}}=T\log_2\left(1+\Gamma_{\mathrm{PR}_{\ell}}^{\mathrm{Dir}}\right)$. In this paper, we thus set $R_{\mathrm{PU}_{\ell},\mathrm{req}}=R_{\mathrm{PT}_{\ell},\mathrm{PR}_{\ell}}$.

We assume all channels experience Rayleigh fading, that randomly distributed as $\sim\mathcal{CN}(0,1)$ and are constant during the duration of the proposed algorithm and subsequent $T$ transmission time-slots. This is a common assumption for auction based and iterative resource allocation algorithms in wireless networks [38–40].

Moreover, the $\mathrm{PTs}$ and $\mathrm{PR}s$ are located on opposite sides of a square of length two, and thus $d_{\mathrm{PT}_\ell,\mathrm{PR}_\ell}{=}2$, while the $\mathrm{STs}$ and $\mathrm{SRs}$ are randomly located in an internal square of length one, located within the square of length two. Moreover, the minimum rate requirements for each $(\mathrm{ST},\mathrm{SR})$ pair is given by $R_{\mathrm{SU,req}}{=}\{R_{\mathrm{SU}_q,\mathrm{req}}\}_{q=1}^{L_{\mathrm{SU}}}{=}0.1$ and all curves are generated based on averaging over 20,000 instances of the algorithm.

Finally, to demonstrate the benefits of the proposed algorithm, we also consider, in addition to the centralized algorithm in (8), a random matching algorithm where the $(\mathrm{PT},\mathrm{PR})$ pairs are randomly matched with the $(\mathrm{ST},\mathrm{SR})$ pairs, and denoted by $\mu^{\mathrm{rnd}}$. When $L_{\mathrm{PU}}{>}L_{\mathrm{SU}}$, $L_{\mathrm{PU}}-L_{\mathrm{SU}}$ $(\mathrm{PT},\mathrm{PR})$ pairs will be randomly omitted from the matching, and when $L_{\mathrm{SU}}{>}L_{\mathrm{PU}}$, $L_{\mathrm{SU}}-L_{\mathrm{PU}}$ $(\mathrm{ST},\mathrm{SR})$ pairs will be randomly omitted from the matching. Once the matchings are randomly established, each matched PT-ST pair start an interactive negotiation process on the values of $\xi_{\ell,q}$ and $\beta_{\ell,q}$. If the PT-ST pair agree on these values, they will cooperate, otherwise there is no cooperation.

We observe that such a random matching requires a centralized approach, and is an upper bound to a completely distributed random matching with no overhead. However, this is sufficient for comparison purposes with the proposed algorithm. Note that the centralized and random matching algorithm represent the two extremes in the amount of overhead and complexity required for any algorithm.

For all figures, unless indicated otherwise the ‘Centralized’ curves are generated by using (8), the ‘Proposed (complete SNR)’ curves are generated as described in Section IV-A1, the ‘Proposed (partial SNR)’ curves are generated as described in Section IV-A2, and the ‘RMBN’ curves are generated based on the random match basic negotiation methods as previously described in this Section.

### *A. Incentive Analysis for Selfish and Rational Users*

A common and realistic assumption in cognitive radio networks is that the primary and secondary users are selfish and rational (see e.g., [1] and [30]). Selfishness implies that users compete with each other to maximize their individual utility function, with no regard for other user’s utility, while rationality implies that users always make decisions to increase their utility. It is thus important for the proposed algorithm to provide, *incentive* for the users to participate in the proposed algorithm, by taking into account the selfish and rational nature of the users.

*1) Accounting for Selfish Users Through Stable Matching:* We first prove that the proposed algorithm takes into consideration the selfish nature of the users, by showing that a stable matching is produced. To define a stable matching[4], we will first define a matching, which is blocked by an individual, and a matching which is blocked by a pair. To do this, we consider $p_\ell{\in}\mathcal{P}$ and $s_q{\in}\mathcal{S}$, where $\mu(p_\ell){\neq}s_q$ and $\mu(s_q){\neq}p_\ell$. A matching $\mu$ is *blocked by an individual* $p_\ell$ $(s_q)$ if $p_\ell$ $(s_q)$ prefers not to be matched, than being matched with its current partner under $\mu$. Mathematically, for $p_\ell$, this implies that $R_{\mathrm{PU}_{\ell,\mu^\dagger(p_\ell)}}(\beta_{\ell,\mu^\dagger(p_\ell)}){<}R_{\mathrm{PU}_{\ell,\mathrm{req}}}$, while for $s_q$, this implies that $R_{\mathrm{SU}_{q,\mu^\dagger(s_q)}}(\beta_{\mu^\dagger(s_q),q}){<}R_{\mathrm{SU}_{q,\mathrm{req}}}$ or $U_{\mathrm{SU}_{q,\mu^\dagger(s_q)}}(\xi_{\mu^\dagger(s_q),q},\beta_{\mu^\dagger(s_q),q}){<}0$.

A matching $\mu$ is *blocked by pair* $(p_\ell, s_q)$ if (i) it is not blocked by individual $p_\ell$ and $s_q$, and (ii) there exists a $\xi_{\ell,q}$ and $\beta_{\ell,q}$ such that $p_\ell$ and $s_q$ can both achieve a higher utility if they were matched together, as opposed to their current matching under $\mu$. The latter condition, thus mathematically implies that

$$\begin{aligned}&U_{\mathrm{PU}_{\ell,\mu^\dagger(p_\ell)}}(\xi_{\ell,\mu^\dagger(p_\ell)},\beta_{\ell,\mu^\dagger(p_\ell)}){<}U_{\mathrm{PU}_{\ell,q}}(\xi_{\ell,q},\beta_{\ell,q})\\&U_{\mathrm{SU}_{q,\mu^\dagger(s_q)}}(\xi_{\mu^\dagger(s_q),q},\beta_{\mu^\dagger(s_q),q}){<}U_{\mathrm{SU}_{q,\ell}}(\xi_{\ell,q},\beta_{\ell,q}).\end{aligned} \tag{15}$$

A matching $\mu$ is then defined as *stable*, under the price and time-slot allocation matrices[5] $\mathbf{G}$ and $\mathbf{B}$, if it is not blocked by any individual or any pair. Given this definition, we present the following theorem.

*Theorem 1:* The proposed algorithm in Section IV-C produces a stable matching, under $\mathbf{G}^{\mathrm{disc}}(\xi_{\mathrm{init}},\delta)$ and $\mathbf{B}^{\mathrm{disc}}(\beta_{\mathrm{init}},\epsilon)$.

*Proof:* See Appendix A. ■

*Theorem 1* is important because it states that the proposed algorithm results in a matching, where any matched $(\mathrm{PT},\mathrm{PR})$ and $(\mathrm{ST},\mathrm{SR})$ pair will not both achieve a higher utility than if they were to respectively partner with any other $(\mathrm{ST},\mathrm{SR})$ and $(\mathrm{PT},\mathrm{PR})$ pair.

*2) Accounting for Rational Users Through Utility Maximization:* It is also important that the proposed algorithm takes into account the rational nature of the users. This first means that each PU and SU will obtain at least a better utility if they participated in the algorithm compared to if they were unmatched. This is equivalent to one of the stable matching conditions; that the matching is not blocked by an individual, and was proved in *Theorem 1*. The second is that at each iteration of the algorithm, each PU and SU should always choose the user, which provides the highest utility. This can be directly observed in MA-Step 2-1 for each PU, and MA-Step 2-2-1-a-i and MA-Step 2-2-1-b for each SU. This implies that none of these matched selfish PUs and SUs have any incentive to deviate from the matching produced by the proposed algorithm.

[4]We extend the common definition of stable matchings [41] to incorporate the price and time-slot allocation numbers

[5]In general, $\mathbf{G}$ and $\mathbf{A}$ can have continuous elements as considered in Section III, or discrete elements as considered in Section IV. Any analysis involving the matching $\mu$ thus has to also consider the domain of the price and time-slot allocation matrices in order to be meaningful.

### B. Utility Performance

We now investigate the utility performance of the proposed algorithm. To do this, we first present the following lemma:

*Lemma 1:* The utility for every $(\mathrm{PT}, \mathrm{PR})$ pair in the stable matching, produced by the proposed algorithm in Section IV-C, is greater than or equal to the utility obtained through a stable matching produced by *any* algorithm, under $\mathbf{G}^{\mathrm{disc}}(\xi_{\mathrm{init}}, \delta)$ and $\mathbf{B}^{\mathrm{disc}}(\beta_{\mathrm{init}}, \epsilon)$.

*Proof:* See Appendix B. ■

*Lemma 1* thus indicates that our algorithm produces the best stable matching, out of every possible stable matchings, under $\mathbf{G}^{\mathrm{disc}}(\xi_{\mathrm{init}}, \delta)$ and $\mathbf{B}^{\mathrm{disc}}(\beta_{\mathrm{init}}, \epsilon)$.

A natural question now arises as to how our algorithm performs when compared with non-stable matchings. To answer this, we first show that our algorithm produces a *weak Pareto optimal* matching, denoted by $\mu_{\mathrm{Pareto}}$. A weak Pareto optimal matching is defined as a matching where there exists at least one matched $(\mathrm{PT}, \mathrm{PR})$ pair under $\mu_{\mathrm{Pareto}}$, which obtains a utility at least greater than any other matching, stable or non-stable. Given this definition, we present the following theorem:

*Theorem 2:* The proposed algorithm in Section IV-C is weak Pareto optimal, under $\mathbf{G}^{\mathrm{disc}}(\xi_{\mathrm{init}}, \delta)$ and $\mathbf{B}^{\mathrm{disc}}(\beta_{\mathrm{init}}, \epsilon)$.

*Proof:* See Appendix C. ■

*Theorem 2* indicates there exists at least one matched $(\mathrm{PT}, \mathrm{PR})$ pair under the proposed algorithm that achieves a utility at least greater than the utility achieved under a non-stable centralized optimal algorithm, under $\mathbf{G}^{\mathrm{disc}}(\xi_{\mathrm{init}}, \delta)$ and $\mathbf{B}^{\mathrm{disc}}(\beta_{\mathrm{init}}, \epsilon)$. This optimal algorithm is similar to the algorithm in (8), but under $\mathbf{G}^{\mathrm{disc}}(\xi_{\mathrm{init}}, \delta)$ and $\mathbf{B}^{\mathrm{disc}}(\beta_{\mathrm{init}}, \epsilon)$, not $\mathbf{G}^{\mathrm{cont}}(\xi_{\mathrm{init}}, \delta)$ and $\mathbf{B}^{\mathrm{cont}}(\beta_{\mathrm{init}}, \epsilon)$.

In fact, the proposed algorithm can achieve a utility for every matched $(\mathrm{PT}, \mathrm{PR})$ pair very close to the centralized optimal algorithm in (8), even under $\mathbf{G}^{\mathrm{cont}}(\xi_{\mathrm{init}}, \delta)$ and $\mathbf{B}^{\mathrm{cont}}(\beta_{\mathrm{init}}, \epsilon)$. This can be observed in Figure 2, which plots the average sum-utility of all matched $(\mathrm{PT}, \mathrm{PR})$ pairs vs. time-slot step number $\epsilon$ for the proposed algorithm, the centralized algorithm in (8), and the RMBN algorithm. Note that the average sum-utility corresponds to the sum over all utilities achieved by the matched $(\mathrm{PT}, \mathrm{PR})$ pairs, averaged over the channel realizations, and given by $U_{\text{PU}_{\Sigma,\mu}}=$ $\sum_{\ell\in\mathcal{P}_\mu} \mathrm{E}\left[U_{\ell,\mu^\dagger(\ell)}(\xi_{\ell,\mu^\dagger(\ell)}, \beta_{\ell,\mu^\dagger(\ell)})\right]$, where $\mathcal{P}_\mu$ corresponds to all the $(\mathrm{PT}, \mathrm{PR})$ pairs matched under $\mu$.

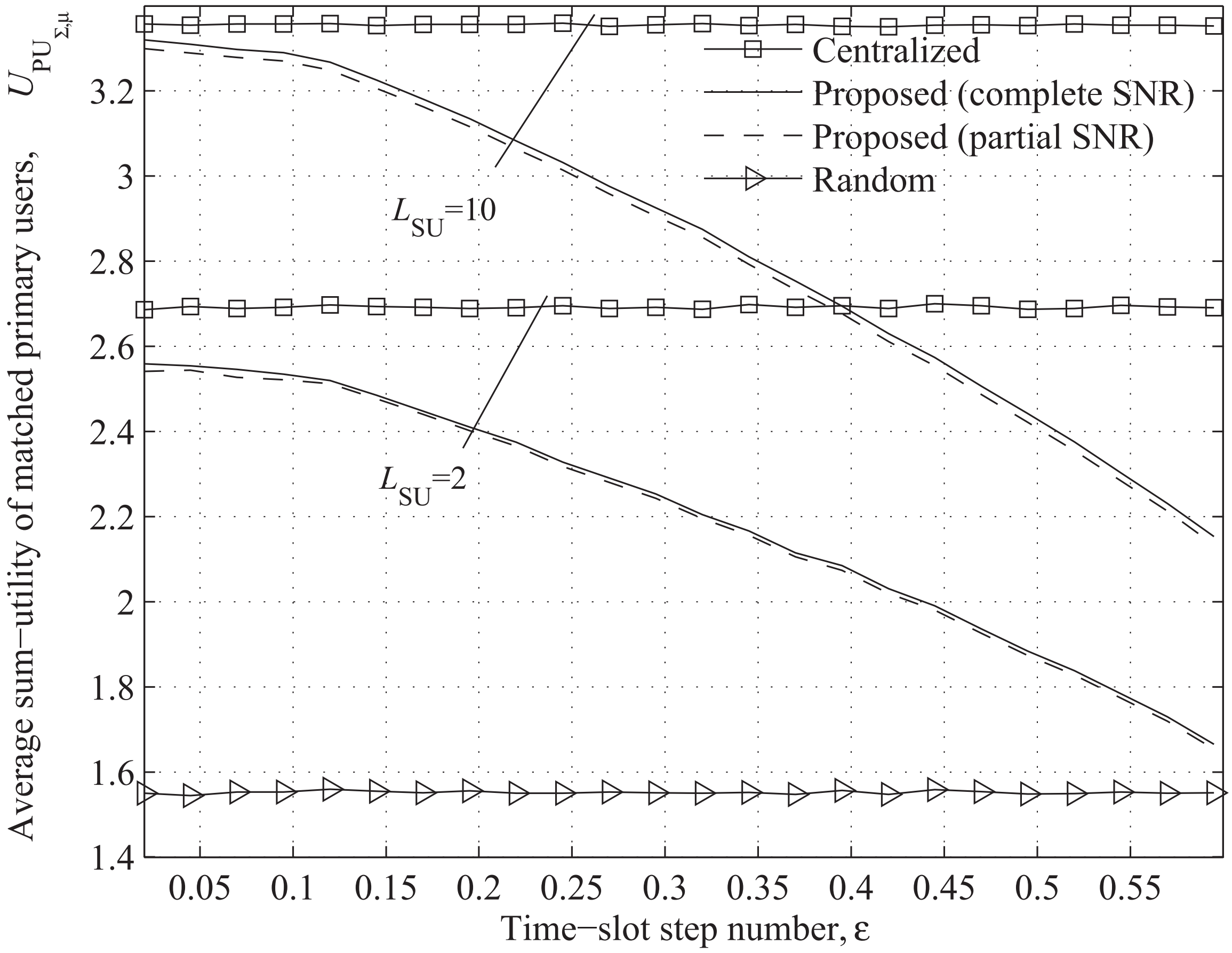

Fig. 2. Average sum-utility of all matched $(\mathrm{PT}, \mathrm{PR})$ pairs vs. time slot step-number $\epsilon$, with $\xi_{\text{init}}$=0.99, $\beta_{\text{init}}$=0.99, $\delta$=$\epsilon$ , $\gamma_{\text{SU}_1}$= $\ldots \gamma_{\text{SU}_{L_{\text{SU}}}}$=25 dB, $L_{\text{PU}}$=2, $\gamma_{\text{PU}_1}$=$\gamma_{\text{PU}_2}$=5 dB, $R_{\text{SU,req}}$=0.1, $\{R_{\text{PU}_\ell,\text{req}}=R_{\text{PT}_\ell,\text{PR}_\ell}\}_{\ell=1}^{L_{\text{PU}}}$, $\bar{c}$=1, $\bar{k}$=1 and $\alpha$=4.

We first observe in Figure 2 that for the proposed algorithm, the complete and partial received SNR scenarios achieve very similar performance, despite different channel assumptions. We next observe that the proposed algorithm (i) achieves a sum-utility comparable with the sum-utility of the centralized algorithm for sufficiently small $\epsilon$, and (ii) performs significantly better than the RMBN algorithm. For example, when the time-slot step number $\epsilon$=0.1 and $L_{\text{SU}}$=10, we observe that the proposed algorithm with complete and partial received SNR achieves respectively (i) $\approx$97% of the sum-utility of the centralized algorithm, and (ii) $\approx$193% sum-utility increase compared to the RMBN algorithm.

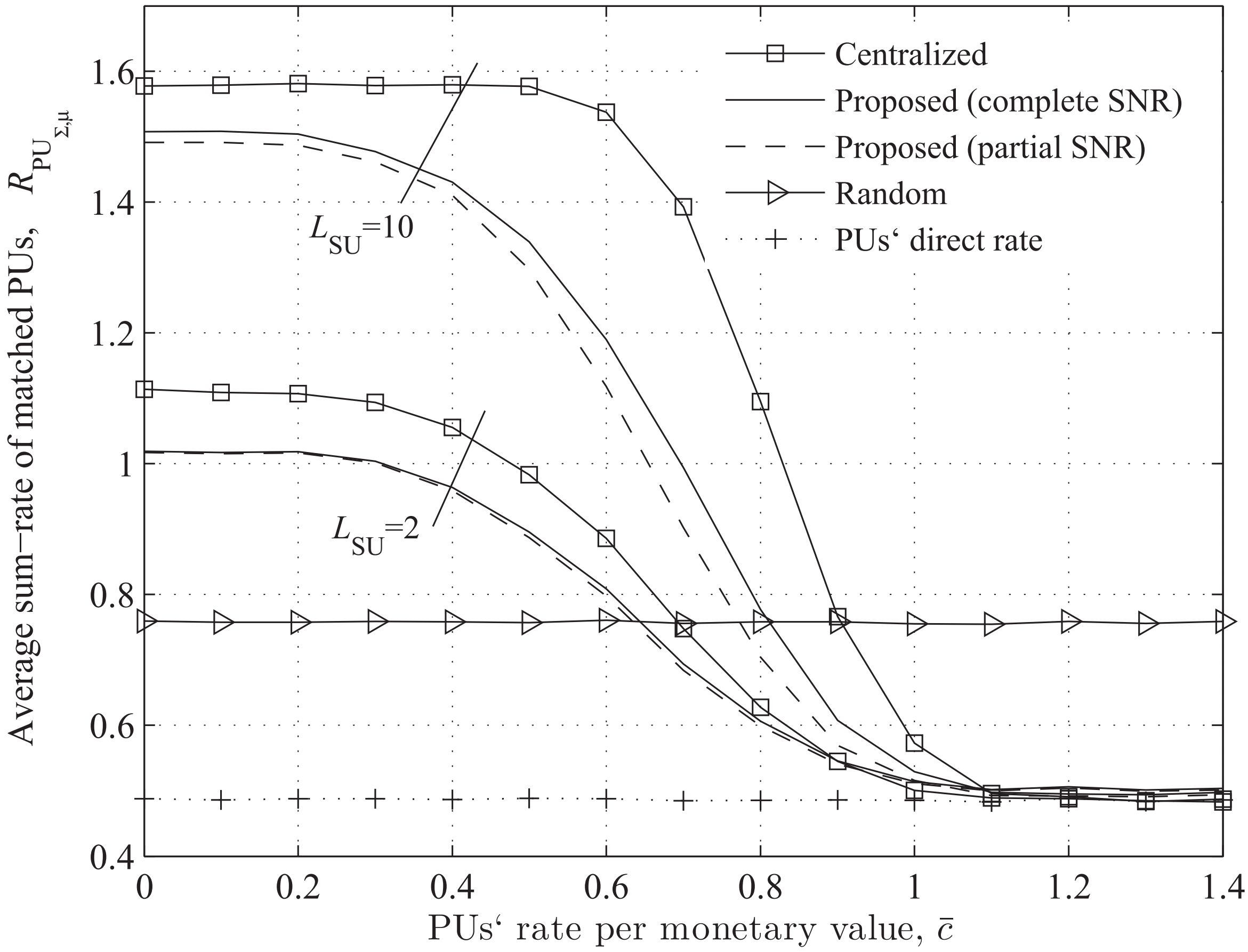

Fig. 3. Average total sum-rate of all matched $(\mathrm{PT}, \mathrm{PR})$ pairs vs. $\bar{c}$, with $\xi_{\text{init}}$=0.99, $\beta_{\text{init}}$=0.99, $\epsilon$=0.05, $\delta$=0.05 , $\gamma_{\text{SU}_1}= \ldots \gamma_{\text{SU}_{L_{\text{SU}}}}$=25 dB, $L_{\text{PU}}$=2, $\gamma_{\text{PU}_1}=\gamma_{\text{PU}_2}$=5 dB, $R_{\text{SU,req}}$=0.1, $\{R_{\text{PU}_\ell,\text{req}}=R_{\text{PT}_\ell,\text{PR}_\ell}\}_{\ell=1}^{L_{\text{PU}}}$, $\bar{k}$=15 and $\alpha$=4.

### C. Rate Performance

As observed in Section IV-C and Figure 2, the proposed algorithm's primary focus is on maximizing the PU's sum-utility, while simultaneously satisfying both the PU's and SU's minimum rate requirement. However, some networks may require that (i) only the rate is important and monetary factors are not a consideration, and (ii) the SUs also prefer a higher rate, greater than their minimum data rate requirement. For such networks, the proposed algorithm is flexible in adapting to these design needs.

In particular, observe from (4) that increasing $\bar{c}$ has the effect of increasing the contribution of the monetary value in the $(\mathrm{PT}, \mathrm{PR})$'s utility function. As observed in the Proposal Update Unit, this has the effect of the PUs choosing to offer a lower time-slot allocation number, rather than a lower price allocation number, to the SUs. This will subsequently result in a higher rate for the SUs, at the expense of a lower rate for the PUs. Thus the parameter $\bar{c}$ can be used to control the rates of the PUs and SUs, in accordance with specific system requirements.

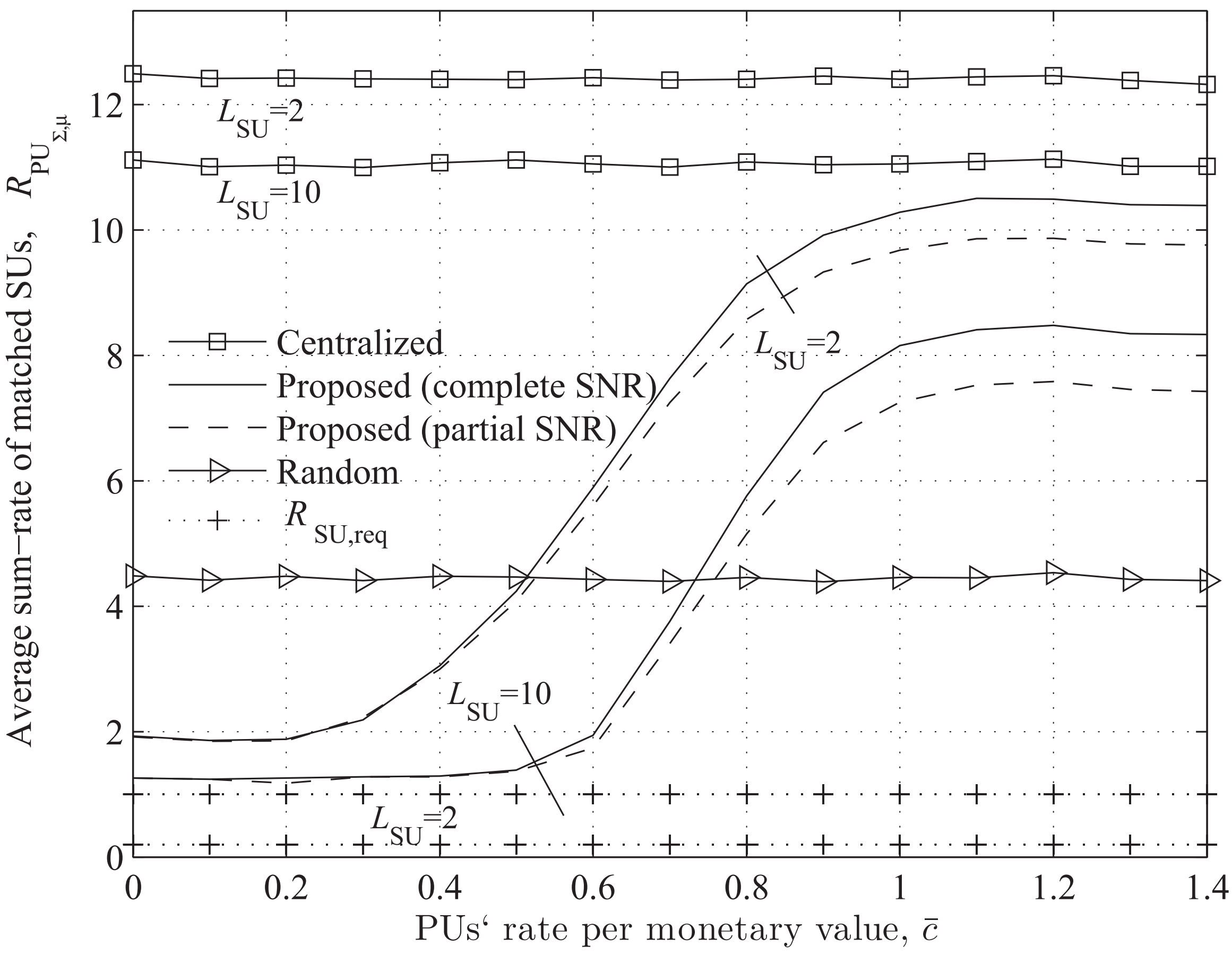

Fig. 4. Average total sum-rate of all matched $(\mathrm{ST},\mathrm{SR})$ pairs vs. $\bar{c}$, with $\xi_{\mathrm{init}}$=0.99, $\beta_{\mathrm{init}}$=0.99, $\epsilon$=0.05, $\delta$=0.05 , $\gamma_{\mathrm{SU}_1}$=$\ldots\gamma_{\mathrm{SU}_{L_{\mathrm{SU}}}}$= 25 dB, $L_{\mathrm{PU}}$=2, $\gamma_{\mathrm{PU}_1}$=$\gamma_{\mathrm{PU}_2}$=5 dB, $R_{\mathrm{SU,req}}$=0.1, $\{R_{\mathrm{PU}_\ell,\mathrm{req}}$=$R_{\mathrm{PT}_\ell,\mathrm{PR}_\ell}\}_{\ell=1}^{L_{\mathrm{PU}}}$, $\bar{k}$=15 and $\alpha$=4.

This can be observed in Figures. 3 and 4, which respectively plots the average sum-rate of all matched $(\mathrm{PT},\mathrm{PR})$ pairs vs. $\bar{c}$, and the average sum-rate of all matched $(\mathrm{ST},\mathrm{SR})$ pairs vs. $\bar{c}$, for the complete received SNR scenario. Note that the average sum-rate of all matched $(\mathrm{PT},\mathrm{PR})$ pairs corresponds to the sum over all rates achieved by the matched $(\mathrm{PT},\mathrm{PR})$ pairs, averaged over the channel realizations, and is given by $R_{\mathrm{PU}_\Sigma,\mu}$=$\sum_{\ell\in\mathcal{P}_\mu}\mathrm{E}\left[R_{\ell,\mu^\dagger(\ell)}(\beta_{\ell,\mu^\dagger})\right]$, where $\mathcal{P}_\mu$ corresponds to all the $(\mathrm{PT},\mathrm{PR})$ pairs matched under $\mu$. A similar definition can be made for the sum-rate of the matched $(\mathrm{ST},\mathrm{SR})$ pairs, denoted $R_{\mathrm{SU}_\Sigma,\mu}$. For comparison, we also plot the sum-rate achieved by a centralized optimal algorithm. In particular, in Figure 3, the centralized algorithm produces a sum-rate, which maximizes the total sum-rate of all $(\mathrm{PT},\mathrm{PR})$ pairs, and given by substituting $\bar{c}$=0 into the optimal algorithm produced from (8). In Figure 4, the centralized algorithm produces a sum-rate which maximizes the sum-rate of all $(\mathrm{ST},\mathrm{SR})$ pairs, and is formulated in a way similar to (8), but interchanging the $(\mathrm{PT},\mathrm{PR})$ pairs with the $(\mathrm{ST},\mathrm{SR})$ pairs in the optimization equation. This results

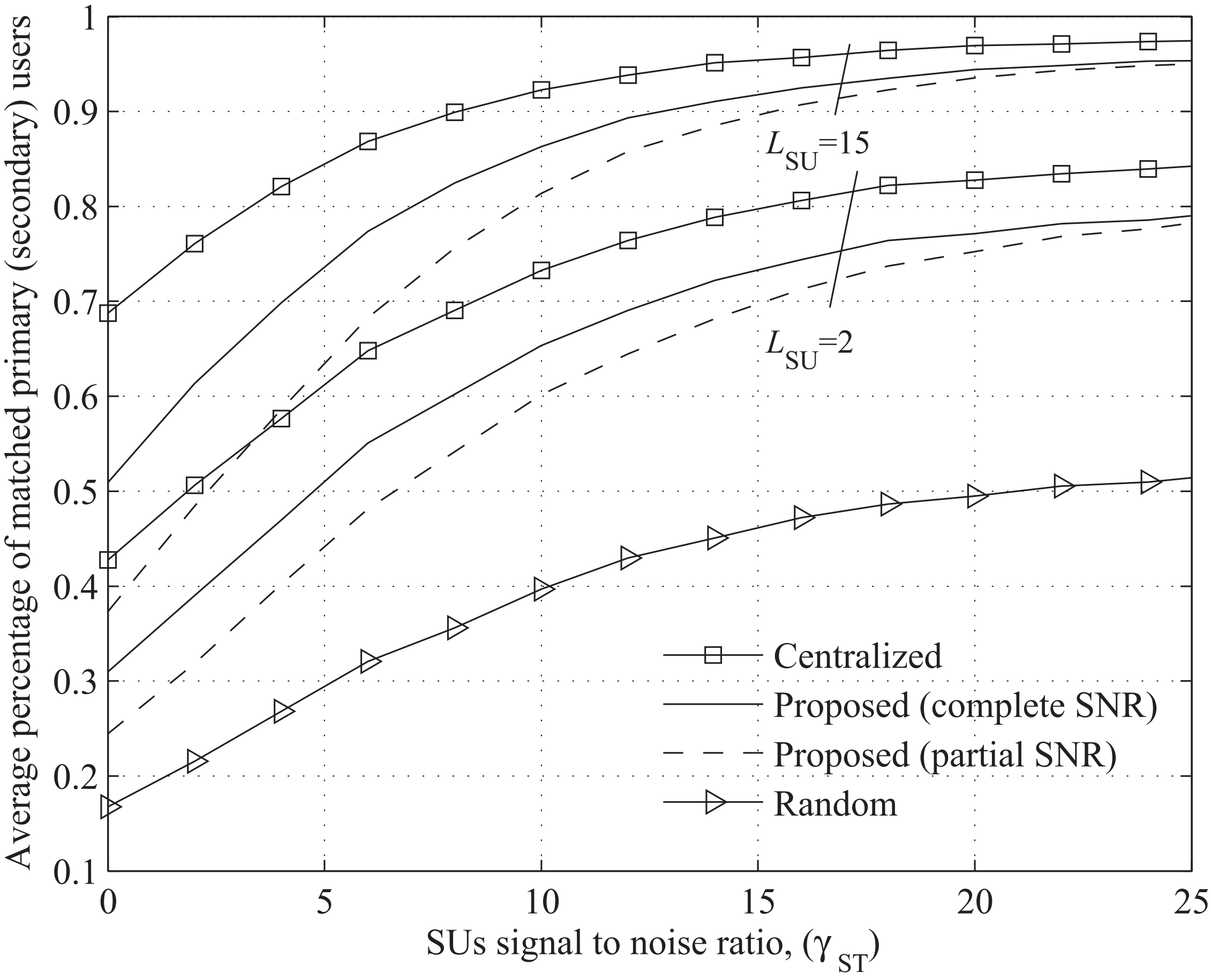

Fig. 5. Percentage of matched $(\mathrm{PT},\mathrm{PR})$ pairs vs. the total number of $(\mathrm{ST},\mathrm{SR})$ pairs $L_{\mathrm{SU}}$, with $\alpha_{\mathrm{init}}$=0.99, $\beta_{\mathrm{init}}$=0.99, $\epsilon$=0.05, $\delta$=0.05 , $\gamma_{\mathrm{SU}_1}$=$\ldots\gamma_{\mathrm{SU}_{L_{\mathrm{SU}}}}$=25 dB, $L_{\mathrm{PU}}$=2, $\gamma_{\mathrm{PU}_1}$=$\gamma_{\mathrm{PU}_2}$=5 dB, $L_{\mathrm{PU}}$=2, $R_{\mathrm{SU,req}}$=0.1, $\{R_{\mathrm{PU}_\ell,\mathrm{req}}=R_{\mathrm{PT}_\ell,\mathrm{PR}_\ell}\}_{\ell=1}^{L_{\mathrm{PU}}}$, $\bar{c}$=1, $\bar{k}$=2 and $\rho$=4.

in the optimization

$$\arg\max_{\{\mathbf{M},\mathbf{B}^{\mathrm{cont}},\mathbf{G}^{\mathrm{cont}}\}}\sum_{\ell=1}^{L_{\mathrm{PU}}}\sum_{q=1}^{L_{\mathrm{SU}}} m_{\ell,q} R_{\mathrm{SU}_{q,\ell}}(\xi_{\ell,q},\beta_{\ell,q})$$

subject to the conditions in (8).

We observe in Figures. 3 and 4 that $R_{\mathrm{PU}_{\Sigma},\mu}$decreases with $\bar{c}$, while $R_{\mathrm{SU}_{\Sigma},\mu}$increases with $\bar{c}$, as expected. This shows the flexibility of our algorithm in adapting to different primary and secondary user priority levels. When $\bar{c}$ is low, we observe in Figure 3 that the sum-rate of the matched $(\mathrm{PT},\mathrm{PR})$ pairs of our proposed algorithm achieves a high percentage of the optimal algorithm and significantly greater than the RMBN algorithm. Similarly, when $\bar{c}$ is high, we observe in Figure 4 that the sum-rate of the matched $(\mathrm{ST},\mathrm{SR})$ pairs of our proposed algorithm achieves a high percentage of the optimal algorithm and significantly greater than the RMBN algorithm.

### D. Total Number of Matchings

The total number of matched $(\mathrm{PT},\mathrm{PR})$ and $(\mathrm{ST},\mathrm{SR})$ pairs is also an important consideration of any matching algorithm, and is proportional to the total number of users which can achieve their minimum rate requirements. Figure 5 shows the percentage of matched $(\mathrm{PT},\mathrm{PR})$ pairs vs. the SU's SNR $\gamma_{\mathrm{ST}}=\gamma_{\mathrm{ST}_1}=\gamma_{\mathrm{ST}_{L_{\mathrm{SU}}}}$, for different number of $(\mathrm{ST},\mathrm{SR})$ pairs $L_{\mathrm{SU}}$. We observe that the percentage of matched $(\mathrm{PT},\mathrm{PR})$ pairs increases with $\gamma_{\mathrm{SU}}$ and $L_{\mathrm{SU}}$, due to the higher achievable rates that the matched $(\mathrm{PT},\mathrm{PR})$ pairs can achieve through cooperative relaying. Remarkably, we observe that the proposed scheme can achieve a very high matching percentage at high SNR even when $L_{\mathrm{SU}}=2$, i.e., $\geq 80\%$ when $\gamma_{\mathrm{SU}}\geq 20$ dB. Moreover, the proposed algorithm delivers a percentage of matched users comparable with the centralized algorithm, and significantly greater than the RMBN algorithm. Note that as the minimum rate requirement for the PUs is equal to the rate of their corresponding direct link transmission without cooperative relaying, Figure 5 thus indicates that the PUs are well motivated to participate in the trading framework with the SUs.

In practice, the unmatched $\mathrm{PT}$s will transmit directly to their corresponding $\mathrm{PR}$s and thus $(\mathrm{PT}_\ell,\mathrm{PR}_\ell)$ will achieve the rate $R_{\mathrm{PT}_\ell,\mathrm{PR}_\ell}$. However, the unmatched $\mathrm{ST}$s will not be able to transmit at all. To remedy this, various modifications to the proposed algorithm can be made, which are the subject of future work such as integrating a fairness mechanism into the algorithm so each $\mathrm{ST}$ has a turn transmitting, though at different times.

### E. Convergence

We will now analyze the convergence behavior of the proposed algorithm, as shown in the following theorem:

*Theorem 3:* The number of iterations required for the proposed algorithm to converge is upper bounded by:

$$\mathcal{I}_{\max}=\frac{\xi_{\mathrm{init}}}{\delta}+\frac{1}{\epsilon}\left(\beta_{\mathrm{init}}-\beta^{\mathrm{MIN}}\right) \tag{16}$$

where $\beta^{\mathrm{MIN}}=\min_{\ell=1,\ldots,L_{\mathrm{PU}}}\min_{q=1,\ldots,L_{\mathrm{SU}}}\beta_{\ell,q}^{\min}$ and $\beta_{\ell,q}^{\min}=\frac{2R_{\mathrm{PU}_{\ell,\mathrm{req}}}}{T\log_2\left(1+\Gamma_{\mathrm{PR}_\ell}^{\mathrm{Dir}}+\Gamma_{\mathrm{PR}_{\ell,q}}^{\mathrm{Relay}}\right)}$.

*Proof:* See Appendix D. ■

*Theorem 3* implies that the proposed algorithm converges after a finite number of iterations, and that this number is dependent on different parameters. For example, we clearly see that the number of iterations for convergence decreases with $\epsilon$ and $\delta$, and increases with $\xi_{\mathrm{init}}$ and $\beta_{\mathrm{init}}$.

### F. Overhead

The proposed algorithm is distributed, and thus incurs significantly less overhead and complexity compared to centralized algorithms. An exact analysis of the amount of overhead and complexity is difficult, due to the dependency on a number of system parameters, such as the minimum sum-rate requirements, the price and time-slot step-numbers and the initial price and time-slot allocation numbers. We can however, find an expression for the upper bound on the maximum number of communication packets between the PTs adn the STs, given in the following theorem:

*Theorem 4:* The number of communication packets between the $\mathrm{PTs}$ and the $\mathrm{STs}$ required in the proposed algorithm is upper bounded by

$$\begin{aligned}\mathcal{N}_{\max}&=(a_1 L_{\mathrm{PU}}+a_2 F)\,\mathcal{I}_{\max} \\ &=(a_1 L_{\mathrm{PU}}+a_2 F)\left(\frac{\xi_{\mathrm{init}}}{\delta}+\frac{1}{\epsilon}\left(\beta_{\mathrm{init}}-\beta^{\mathrm{MIN}}\right)\right)\end{aligned} \tag{17}$$

where where $a_1, a_2 \in \mathbb{R}^+$ and $F=\max\{L_{\mathrm{PU}}, L_{\mathrm{SU}}\}$.

*Proof:* See Appendix E. ■

We observe in (17) that the amount of overhead, and thus the number of iterations, decreases with $\epsilon$ and $\delta$. This is confirmed in Figure 6, which plots the total number of communication packets exchanged between each $\mathrm{PT}$ and all the $\mathrm{STs}$ it communicates with, vs. time-slot step number $\epsilon$, with the same parameters used in Figure 2. We see that the total number of communication packets converge to a constant at sufficiently high $\epsilon$. This is because if $\epsilon$ is sufficiently large, the time-slot allocation numbers are updated in the algorithm in such a way that the preference lists for each $(\mathrm{PT}, \mathrm{PR})$ and $(\mathrm{ST}, \mathrm{SR})$ pair remain unchanged.

Denoting the random variable $Y$ as the total number of communication packets as a function of the user channels, Figure 7 plots the cumulative distribution function (c.d.f.) of $Y$, $\mathrm{PR}\,(Y \leq y)$, vs. $y$ for different number of secondary users $L_{\mathrm{SU}}$, where $y$ is a realization of $Y$. We observe that the proposed algorithm can achieve low overhead with high probability in various practical scenarios. For example, when $L_{\mathrm{SU}}=6$, we observe that $90\%$ of the time, a maximum of only $15$ communication packets are exchanged between the PUs and the SUs.

Figures. 2 and 6, and (17) reveal that $\epsilon$ can be designed to ensure an acceptable amount of overhead and achievable rate. In particular, we observe from (17) and Figure 6 that decreasing the overhead by increasing $\epsilon$ or $\delta$ will result in both a lower magnitude and number of price and time-slot allocation numbers which can be offered to the $(\mathrm{ST},\mathrm{SR})$ pairs from the $(\mathrm{PT},\mathrm{PR})$ pairs. This will result in a lower utility for the $(\mathrm{PT},\mathrm{PR})$ pairs, and increase the chance the STs will reject any offer made. This can be observed in Figure 2, which illustrates a tradeoff between performance and overhead. A similar argument can be made for decreasing $\xi_{\text{init}}$ and $\beta_{\text{init}}$. In practice, each PU can adaptively adjust it's time-slot step number $\epsilon$ based on (i) the acceptable sum utility of PUs in Figure 2 and (ii) the tolerable average number of required communication packets in Figure 6.

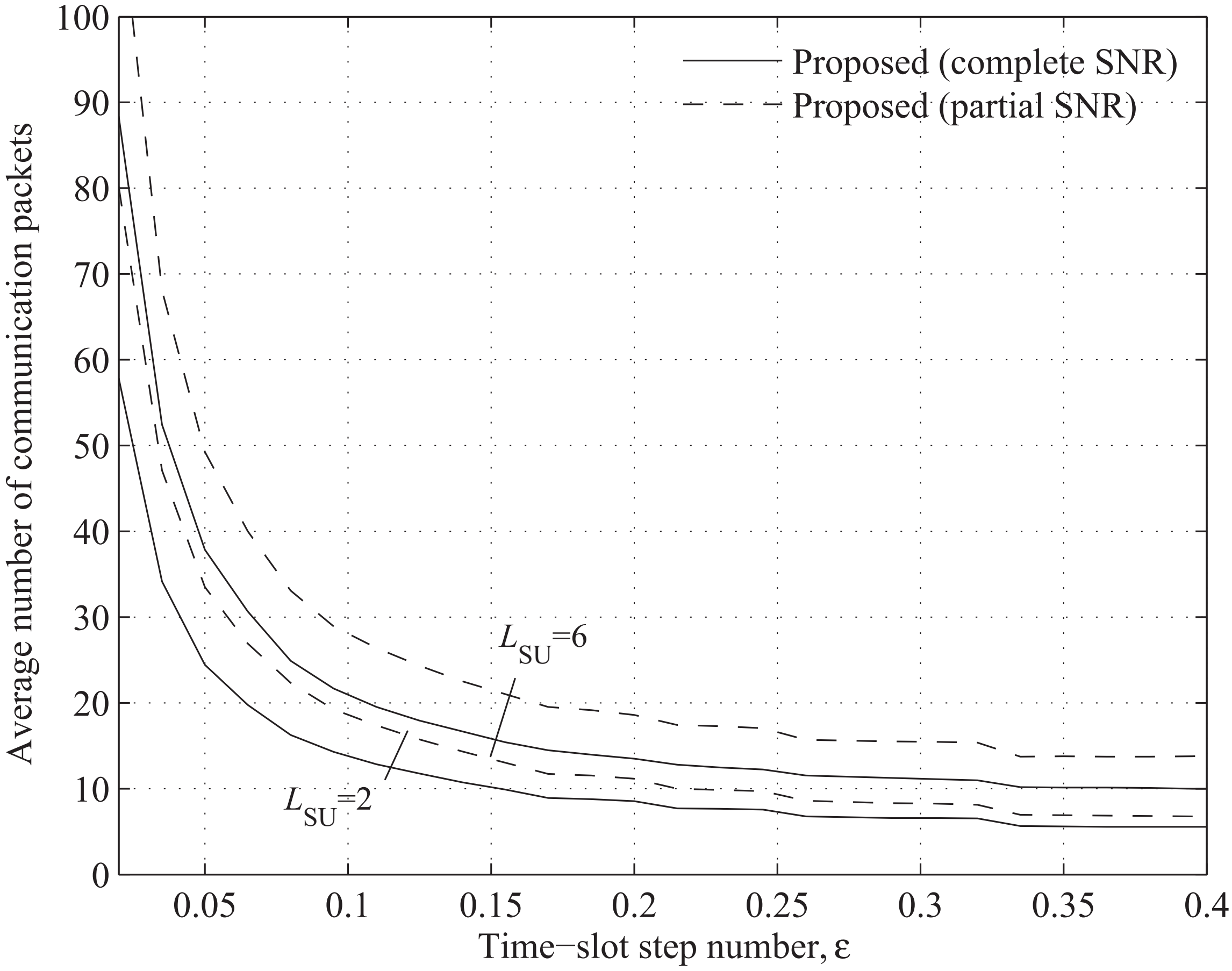

Fig. 6. Total number of communication packets vs. $\epsilon$ for the complete instantaneous received SNR scenario, with $\alpha_{\text{init}}$=0.99, $\beta_{\text{init}}$=0.99, $\epsilon$=0.05, $\delta$=0.05 , $\gamma_{\text{SU}_1}=\dots\gamma_{\text{SU}_{L_{\text{SU}}}}$=25 dB, $L_{\text{PU}}$=2, $\gamma_{\text{PU}_1}=\gamma_{\text{PU}_2}$=5 dB, $R_{\text{SU,req}}$=0.1, $\{R_{\text{PU}_\ell,\text{req}}=R_{\text{PT}_\ell,\text{PR}_\ell}\}_{\ell=1}^{L_{\text{PU}}}$, $\bar{c}$=1, $\bar{k}$=1 and $\rho$=4.

We note that the packet length required for communication between the $\mathrm{PT}$s and the $\mathrm{ST}$s is very short. In particular, assuming that $\xi_{\text{init}}$, $\beta_{\text{init}}$, $\delta$ and $\epsilon$ are initially known to all users, each $\mathrm{PT}$ is only required to send *one bit* to the first $\mathrm{ST}$ in its preference list indicating an offer, and the corresponding $\mathrm{ST}$ only needs to send *one bit* back to the offering $\mathrm{PT}$ indicating either acceptance or rejection. The required time for the proposed algorithm execution is thus very small. Specifically, if high speed control channels such as 802.11 are used, the average time for execution is in the order of 100-200 nsec [42]. As demonstrated in Figure 6, the total number of communication packets for each $\mathrm{PT}$ can be designed to be reasonably small, and thus given the short packet lengths, the total running time and amount of overhead from the proposed algorithm can be quite small.

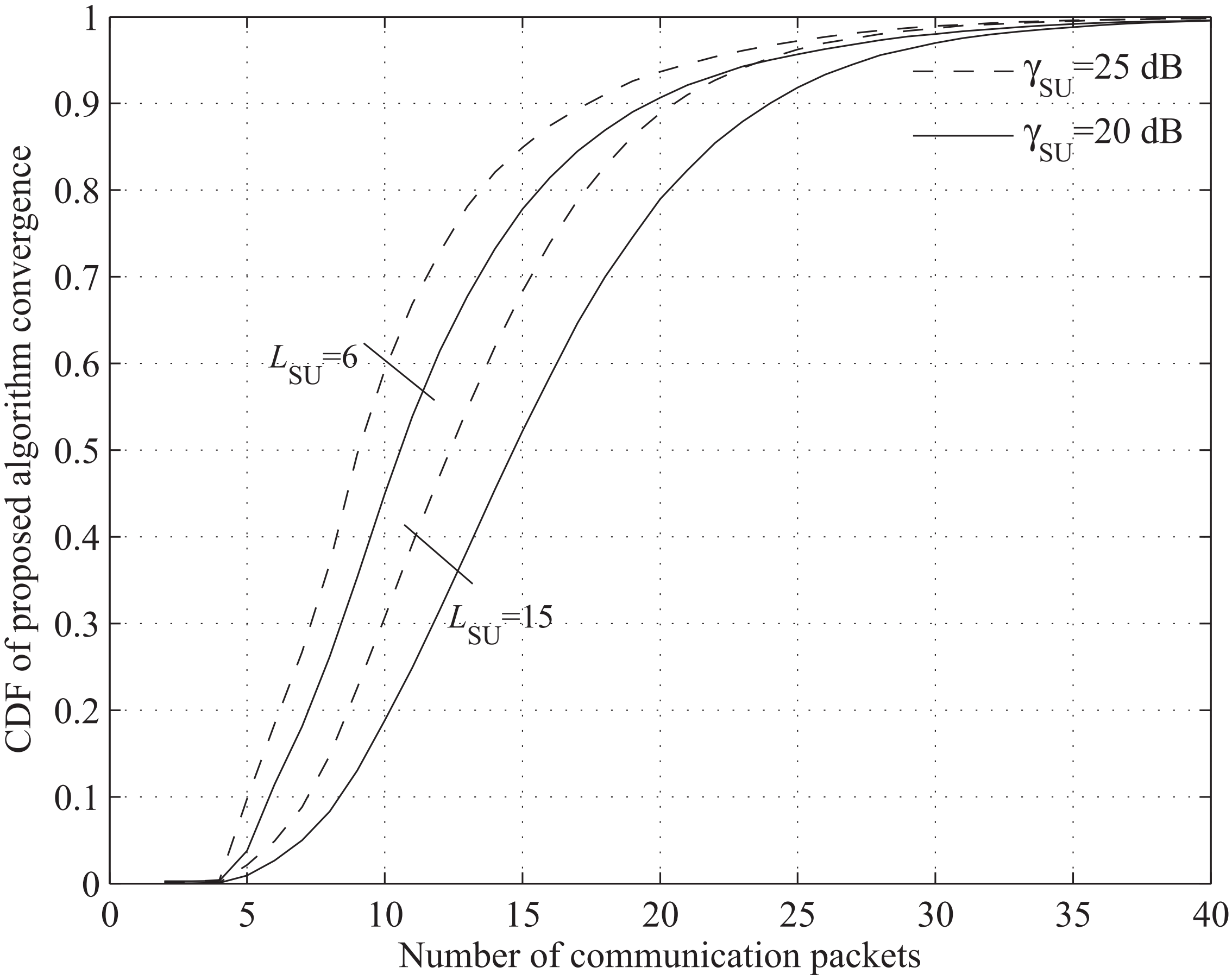

Fig. 7. Cumulative distribution function vs. total number of communication packets for the complete instantaneous received SNR scenario, with $\alpha_{\rm init}$=0.99, $\beta_{\rm init}$=0.99, $\epsilon$=0.05, $\delta$=0.05 , $\gamma_{{\rm SU}_1}$=$\ldots\gamma_{{\rm SU}_{L_{\rm SU}}}$=25 dB, $L_{\rm PU}$=2, $\gamma_{{\rm PU}_1}$=$\gamma_{{\rm PU}_2}$=5 dB, $R_{\rm SU,req}$=0.1, $\{R_{{\rm PU}_\ell,{\rm req}}$=$R_{{\rm PT}_\ell,{\rm PR}_\ell}\}_{\ell=1}^{L_{\rm PU}}$, $\bar{c}$=1, $\bar{k}$=1 and $\rho$=4.

## *G. Complexity*

We now present a lemma for the complexity of the centralized, proposed and RMBN algorithms:

*Lemma 2:* The complexity of the centralized algorithm is given by

$$\begin{cases} \mathcal{O}\left(\frac{L_{\rm SU}!}{(L_{\rm SU}-L_{\rm PU})!}2^{2L_{\rm SU}+L_{\rm PU}}\right), & \text{if } L_{\rm SU}\geq L_{\rm PU}; \\ \mathcal{O}\left(\frac{L_{\rm PU}!}{(L_{\rm PU}-L_{\rm SU})!}2^{2L_{\rm SU}+L_{\rm PU}}\right), & \text{if } L_{\rm SU}<L_{\rm PU}, \end{cases} \tag{18}$$

of the proposed algorithm by $\mathcal{O}\left(L_{\rm PU}L_{\rm SU}\right)$, and of the RMBN algorithm by $\mathcal{O}\left(L_{\rm PU}\right)$.

*Proof:* See Appendix F. ■

We observe that the centralized method has a significantly larger complexity than both the proposed and randomized algorithm. Its complexity is increasing exponentially with the number of primary and secondary users. In contrast, the proposed algorithm complexity only increases linearly with the number of primary or secondary users.

## VI. Conclusion

We proposed a distributed algorithm for spectrum access, which guarantees that the PUs' and SUs' rate requirements are satisfied. Our algorithm hinges on a trading framework between the PUs and SUs, where the PUs and SUs negotiate on combined time-slot and monetary compensation. Time slot allocation numbers determine the amount of transmission time dedicated for SU transmission, and the length of time the SUs will cooperative relay the PUs' data. The price allocation numbers express the amount of monetary compensation that SU provide for PUs in exchange of spectrum usage. The proposed algorithm, was based on a dynamic bilateral negotiation between the PUs and SUs that resulted in a stable outcome. We proved that the proposed algorithm results in the best possible stable matching and is weak Pareto optimal. A numerical analysis also revealed that the distributed algorithm achieves a performance comparable to an optimal centralized algorithm, but with significantly less overhead and complexity.

## Appendix

### A. Proof of Theorem 1

We first define some notations. Without loss of generality, let $\{(\mathrm{PT}_\ell \mathrm{PR}_\ell)\}_{\ell=1}^{L_{\mu^{\mathrm{prop}}}}$ be the set of matched $(\mathrm{PT}, \mathrm{PR})$ pairs at the conclusion of the algorithm. We first consider $p_1{=}(\mathrm{PT}_1, \mathrm{PR}_1)$, with the preference list at the conclusion of the algorithm denoted by $\overline{\mathrm{PULIST}}_1{=}\{s_1, \ldots, s_{K_1}\}$. Associated with this preference list are the final price and time-slot allocation numbers associating $p_1$ and $\{s_q\}_{q=1}^{K_1}$, and denoted respectively by $\{\xi_{1,q}^{\mathrm{prop}}\}_{q=1}^{K_1}$ and $\{\beta_{1,q}^{\mathrm{prop}}\}_{q=1}^{K_1}$. Note that due to the ordering of the preference list, the proposed algorithm will match $p_1$ with $s_1$. We now prove that the proposed algorithm doesn't produce any blocking pairs by contradiction. Assume that $p_1$ and $s_q$, for some $q{=}2, \ldots, K_1$, constitute a blocking pair. This implies that there exists a $\xi_{1,q}$ and $\beta_{1,q}$ such that $U_{\mathrm{PU}_{1,q}}(\xi_{1,q}, \beta_{1,q}){>}U_{\mathrm{PU}_{1,1}}(\xi_{1,1}^{\mathrm{prop}}, \beta_{1,1}^{\mathrm{prop}})$, implying that either (i) $\xi_{1,q}{>}\xi_{1,q}^{\mathrm{prop}}$, (ii) $\beta_{1,q}{>}\beta_{1,q}^{\mathrm{prop}}$ or (iii) both $\xi_{1,q}{>}\xi_{1,q}^{\mathrm{prop}}$ and $\beta_{1,q}{>}\beta_{1,q}^{\mathrm{prop}}$. Due to the ordering of the preference list $\mathrm{PULIST}_1$, this cannot occur as there would have been a period during the algorithm when $s_q$ initially rejected $p_1$ at this $\beta_{1,q}$ and/or $\xi_{1,q}$ value, as $s_q$ received a better offer from another $(\mathrm{PT}, \mathrm{PR})$ pair. For this scenario, $p_1$ and $s_q$ thus do not form a blocking pair. A similar proof can be made for the other $p_\ell$s, for $\ell{=}2, \ldots, L_{\mathrm{PU}}$. The stable matching proof follows by noting that there are no blocking individuals due to the preference lists, i.e., no $(\mathrm{PT},\mathrm{PR})$ and $(\mathrm{ST},\mathrm{SR})$ pairs will be matched respectively with a $(\mathrm{ST},\mathrm{SR})$ and $(\mathrm{PT},\mathrm{PR})$ pair not on its preference list.

### *B. Proof of Lemma 1*

The proof is by contradiction. Denote $\mu^{\mathrm{prop}}$ as the stable matching produced by the proposed algorithm, and $\mu^{\mathrm{alt}}$ as another stable matching. Then let us assume that for all $(\mathrm{PT},\mathrm{PR})$ pairs, the utility achieved under $\mu^{\mathrm{prop}}$ is less than the utility achieved under $\mu^{\mathrm{alt}}$. W.l.o.g, consider $(\mathrm{PT}_\ell,\mathrm{PR}_\ell)$, and thus the previous statement mathematically implies that $U_{\mathrm{PU}_{\ell,\mu^{\dagger\mathrm{alt}}(\ell)}}(\xi^{\mathrm{alt}}_{\ell,\mu^{\dagger\mathrm{alt}}(\ell)},\beta^{\mathrm{alt}}_{\ell,\mu^{\dagger\mathrm{alt}}(\ell)})>U_{\mathrm{PU}_{\ell,\mu^{\dagger\mathrm{prop}}(\ell)}}(\xi^{\mathrm{prop}}_{\ell,\mu^{\dagger\mathrm{prop}}(\ell)},\beta^{\mathrm{prop}}_{\ell,\mu^{\dagger\mathrm{prop}}(\ell)})$, where $(\xi^{\mathrm{prop}}_{\cdot,\cdot},\beta^{\mathrm{prop}}_{\cdot,\cdot})$ and $(\xi^{\mathrm{alt}}_{\cdot,\cdot},\beta^{\mathrm{alt}}_{\cdot,\cdot})$ denote the price and time-slot allocation numbers under matching $\mu^{\mathrm{prop}}$ and $\mu^{\mathrm{alt}}$ respectively. Denoting $q{=}\mu^{\dagger\mathrm{alt}}(\ell)$, this means that under the proposed matching $\mu^{\mathrm{prop}}$, $\mathrm{PT}_\ell$ offered $(\xi^{\mathrm{alt}}_{\ell,\mu^{\dagger\mathrm{alt}}(\ell)},\beta^{\mathrm{alt}}_{\ell,\mu^{\dagger\mathrm{alt}}(\ell)})$ to $\mathrm{ST}_q$ but was rejected. There are two ways for this to happen. The first is if under matching $\mu^{\mathrm{prop}}$, $\mathrm{PT}_\ell$ with proposal $(\xi^{\mathrm{alt}}_{\ell,\mu^{\dagger\mathrm{alt}}(\ell)},\beta^{\mathrm{alt}}_{\ell,\mu^{\dagger\mathrm{alt}}(\ell)})$ was not in $\mathrm{ST}_q$'s preference list $\mathrm{SULIST}_q$. However, this contradicts the implicit assumption that under matching $\mu^{\mathrm{alt}}$, $\mathrm{PT}_\ell$ with proposal $(\xi^{\mathrm{alt}}_{\ell,\mu^{\dagger\mathrm{alt}}(\ell)},\beta^{\mathrm{alt}}_{\ell,\mu^{\dagger\mathrm{alt}}(\ell)})$ is in $\mathrm{ST}_q$'s preference list, and thus should also be in $\mathrm{ST}_q$'s preference list under matching $\mu^{\mathrm{prop}}$. The second alternative is if under matching $\mu^{\mathrm{prop}}$, $\mathrm{ST}_q$ rejected $\mathrm{PT}_\ell$ in favor of another primary transmitter, denoted as $\mathrm{PT}_\nu$. W.l.o.g., let us assume that this is the first rejection that has taken place under $\mu^{\mathrm{prop}}$. This implies that $\mathrm{PT}_\nu$ prefers $\mathrm{ST}_q$ to every other $(\mathrm{ST},\mathrm{SR})$ pair, and thus the matching $\mu^{\mathrm{alt}}$ is blocked by pair $((\mathrm{PT}_\nu,\mathrm{PR}_\nu),(\mathrm{ST}_q,\mathrm{SR}_q))$, which contradicts the assumption that $\mu^{\mathrm{alt}}$ is a stable matching.

### *C. Proof of Theorem 2*

We outline the proof for non-stable matchings, and note that the proof for stable matchings follows directly from Lemma 1. The proof is by contradiction. Denote $\mu^{\mathrm{prop}}$ as the matching produced by the proposed algorithm, and $\mu^{\mathrm{opt}}$ as an arbitrary non-stable matching. Then let us assume that every matched $(\mathrm{PT},\mathrm{PR})$ pair in $\mu^{\mathrm{prop}}$ achieves a utility less than its utility obtained under matching $\mu^{\mathrm{opt}}$. Mathematically, this is represented by $\{U_{\ell,\mu^{\dagger\mathrm{prop}}(\ell)}(\xi^{\mathrm{prop}}_{\ell,\mu^{\dagger\mathrm{prop}}(\ell)},\beta^{\mathrm{prop}}_{\ell,\mu^{\dagger\mathrm{prop}}(\ell)})<U_{\ell,\mu^{\dagger\mathrm{opt}}(\ell)}(\xi^{\mathrm{opt}}_{\ell,\mu^{\dagger\mathrm{opt}}(\ell)},\beta^{\mathrm{opt}}_{\ell,\mu^{\dagger\mathrm{opt}}(\ell)})\}_{\ell=1}^{L_{\mu^{\mathrm{prop}}}}$, where $(\xi^{\mathrm{prop}}_{\cdot,\cdot},\beta^{\mathrm{prop}}_{\cdot,\cdot})$ and $(\xi^{\mathrm{opt}}_{\cdot,\cdot},\beta^{\mathrm{opt}}_{\cdot,\cdot})$ denote the price and time-slot allocation numbers under matching $\mu^{\mathrm{prop}}$ and $\mu^{\mathrm{opt}}$ respectively. We consider three scenarios, defined by the relative number of matched $(\mathrm{ST},\mathrm{SR})$ pairs under $\mu^{\mathrm{prop}}$ compared to $\mu^{\mathrm{opt}}$. Specifically:

*1) Same matchings:* In this scenario, every matched $(\mathrm{ST},\mathrm{SR})$ and $(\mathrm{PT},\mathrm{PR})$ pair under $\mu^{\mathrm{prop}}$ is also matched under $\mu^{\mathrm{opt}}$. Consider the *last* $(\mathrm{ST},\mathrm{SR})$ pair which is matched in $\mu^{\mathrm{prop}}$, and denote this pair as $(\mathrm{ST}_{L_{\mu\mathrm{prop}}},\mathrm{SR}_{L_{\mu\mathrm{prop}}})$. Moreover, w.l.o.g. $(\mathrm{ST}_{L_{\mu\mathrm{prop}}},\mathrm{SR}_{L_{\mu\mathrm{prop}}})$ is matched with (i) $(\mathrm{PT}_{L_{\mu\mathrm{prop}}},\mathrm{PR}_{L^{\mu\mathrm{prop}}})$ under matching $\mu^{\mathrm{prop}}$, and (ii) $(\mathrm{PT}_1,\mathrm{PR}_1)$ under matching $\mu^{\mathrm{opt}}$. From the initial contradiction assumption, note that under $\mu^{\mathrm{prop}}$, $(\mathrm{PT}_1,\mathrm{PR}_1)$ prefers $(\mathrm{ST}_{L_{\mu\mathrm{prop}}},\mathrm{SR}_{L_{\mu\mathrm{prop}}})$ with price and time slot allocation numbers $(\xi^{\mathrm{alt}}_{1,L_{\mu\mathrm{prop}}},\beta^{\mathrm{alt}}_{1,L_{\mu\mathrm{prop}}})$ than its current matching, which mathematically implies that

$U_{\mathrm{PU}_{1,L_{\mu\mathrm{prop}}}}(\xi^{\mathrm{alt}}_{1,L_{\mu\mathrm{prop}}},\beta^{\mathrm{alt}}_{1,L_{\mu\mathrm{prop}}})>U_{\mathrm{PU}_{1,\mu^{\dagger\mathrm{prop}}(1)}}(\xi^{\mathrm{prop}}_{1,\mu^{\dagger\mathrm{prop}}(1)},\beta^{\mathrm{prop}}_{1,\mu^{\dagger\mathrm{prop}}(1)})$. There was thus a period during the proposed algorithm where $(\mathrm{PT}_1,\mathrm{PR}_1)$ offered a price allocation number $\xi^{\mathrm{alt}}_{1,L_{\mathrm{PU}}}(>\xi^{\mathrm{prop}}_{1,L_{\mathrm{PU}}})$, and/or a time-slot allocation number $\beta^{\mathrm{alt}}_{1,L_{\mathrm{PU}}}(>\beta^{\mathrm{prop}}_{1,L_{\mathrm{PU}}})$ to $(\mathrm{ST}_{L_{\mu\mathrm{prop}}},\mathrm{SR}_{L_{\mu\mathrm{prop}}})$, and was rejected. However, as $(\mathrm{ST}_{L_{\mu\mathrm{prop}}},\mathrm{SR}_{L_{\mu\mathrm{prop}}})$ is the last $(\mathrm{ST},\mathrm{SR})$ pair to be matched, this is a contradiction as $(\mathrm{ST}_{L_{\mu\mathrm{prop}}},\mathrm{SR}_{L_{\mu\mathrm{prop}}})$ will not reject any offer from $(\mathrm{PT}_1,\mathrm{PR}_1)$.

*2) More primary user matchings:* In this scenario, there are more $(\mathrm{PT},\mathrm{PR})$ pairs which are matched under $\mu^{\mathrm{prop}}$, then matched under $\mu^{\mathrm{opt}}$. The initial contradiction assumption is thus clearly violated, as the un-matched $(\mathrm{PT},\mathrm{PR})$ pairs which are matched under $\mu^{\mathrm{prop}}$, but not under $\mu^{\mathrm{opt}}$, have a zero utility.

*3) Less primary user matchings:* In this scenario, there are less $(\mathrm{PT},\mathrm{PR})$ pairs which are matched under $\mu^{\mathrm{prop}}$, then matched under $\mu^{\mathrm{opt}}$. Every $(\mathrm{ST},\mathrm{SR})$ pair which is not matched under $\mu_{\mathrm{prop}}$ will thus have a zero utility. However, at least one of these $(\mathrm{ST},\mathrm{SR})$ pairs, denoted as $(\mathrm{ST}_q,\mathrm{SR}_q)$ will obtain a positive utility under $\mu_{\mathrm{opt}}$. Combined with the initial contradiction assumption, this implies that $(\mathrm{ST}_q,\mathrm{SR}_q)$ and $(\mathrm{PT}_{\mu^{\dagger\mathrm{opt}}(q)},\mathrm{PR}_{\mu^{\dagger\mathrm{opt}}(q)})$ form a blocking pair for $\mu^{\mathrm{prop}}$, which is a contradiction.

## D. Proof of Theorem 3

From MA-Step 2-2-1-a-i and MA-Step 2-2-2 of the proposed algorithm in Table II, if $\mathrm{ST}_q$ rejects the offer of $\mathrm{PT}_\ell$, then $\mathrm{PT}_\ell$ will decrease its price allocation number $\beta_{\ell,q}$ or its time-slot allocation number $\xi_{\ell,q}$ by executing PUU. The minimum possible price allocation number that $\mathrm{PT}_\ell$ can offer $\mathrm{ST}_q$ is $\xi_{\ell,q}{=}0$, while the minimum time-slot allocation number that $\mathrm{PT}_\ell$ can offer $\mathrm{ST}_q$ is

$$\beta^{\min}_{\ell,q}{=}\frac{2R_{\mathrm{PU}_{\ell,\mathrm{req}}}}{T\log_2\left(1+\Gamma^{\mathrm{Dir}}_{\mathrm{PR}_\ell}+\Gamma^{\mathrm{Relay}}_{\mathrm{PR}_{\ell,q}}\right)}. \tag{19}$$

$\mathrm{PT}_\ell$ will not decrease its time-slot allocation number to $\beta_{\ell,q}{=}\beta^{\min}_{\ell,q}-\epsilon$, since its minimum rate requirement will not be satisfied. Thus the minimum possible time-slot allocation number that $\mathrm{PT}_\ell$ can offer in the algorithm is

$$\beta_{\ell}^{\mathrm{MIN}}=\min_{q=1,\ldots,L_{\mathrm{SU}}}\beta_{\ell,q}^{\min}. \tag{20}$$

So in the worst case scenario, each $\mathrm{PT}_{\ell}$, $\ell=1,\ldots,L_{\mathrm{PU}}$ after $\lceil\frac{\xi_{\mathrm{init}}}{\delta}+\frac{1}{\epsilon}\left(\beta_{\mathrm{init}}-\beta_{\ell}^{\mathrm{MIN}}\right)\rceil$ iterations will not make any offer to any ST, and according to MA Step 2-2-3, $\mathrm{PT}_{\ell}$ will be removed from the list of unmatched PTs denoted by $\mathrm{MATCH}$ in Table II. Thus the maximum number of iterations that all the $\mathrm{PTs}$ will be removed from $\mathrm{MATCH}$ is given by

$$\mathcal{I}_{\max}=\frac{\xi_{\mathrm{init}}}{\delta}+\frac{1}{\epsilon}\left(\beta_{\mathrm{init}}-\min_{\ell=1,\ldots,L_{\mathrm{PU}}}\beta_{\ell}^{\mathrm{MIN}}\right). \tag{21}$$

*E. Proof of Theorem 4*

Following the proposed algorithm in Table II, the $\mathrm{PTs}$ and $\mathrm{STs}$ need to communicate together as follows. In MA-Step 2-1, in the worst case scenario, all the $\mathrm{PTs}$ make an offer to all the $\mathrm{STs}$, and thus $L_{\mathrm{PU}}$ communication packets are required. After receiving the PTs offer, the $\mathrm{STs}$ need to inform the PTs whether they accepted or rejected the offers. This process happens in MA-Step 2-2-1-a-i, MA-Step 2-2-1-a-ii and MA-Step 2-2-2 and thus $F$ communication packets are required. Thus in the first iteration of the algorithm, a maximum of $L_{\mathrm{PU}}+F$ communication packets are required.

However, since in the next iterations some of the PUs are matched and also the length of the preference list of the unmatched PTs is decreased, the number of communication packets will decrease. Thus after the first iteration, the total number of communication packets at each iteration of the algorithm is a linear function of $L_{\mathrm{PU}}$ and $F$. The maximum number of iterations that the algorithm ends is also known from Theorem 3. Therefore, the maximum number of communication packets packets between the $\mathrm{PTs}$ and $\mathrm{STs}$ scales as $(a_1 L_{\mathrm{PU}}+a_2 F)\,\mathcal{I}_{\max}$.

*F. Proof of Lemma 2*

For the centralized method when $L_{\mathrm{PU}}\leqslant L_{\mathrm{SU}}$, the proof follows by noting that the total number of matching combinations between the PTs and STs is $\frac{L_{\mathrm{SU}}!}{(L_{\mathrm{SU}}-L_{\mathrm{PU}})!}$, while the complexity of solving the linear programming problem for all possible matching combinations is $2^{2L_{\mathrm{SU}}+L_{\mathrm{PU}}}$ [43]. A similar argument can be made when $L_{\mathrm{PU}}>L_{\mathrm{SU}}$. For the proposed algorithm, the proof follows by noting that the complexity is proportional to the total number of times the PTs communicates with the STs, given in (17). For the RMBN method, the proof follows by noting that each PT communicates to a random ST only once, and thus the complexity is proportional to the number of primary users $L_{\mathrm{PU}}$.

## ACKNOWLEDGMENT

This work was supported in part by the Australian Research Council (ARC) Discovery Projects Grants DP120100190, DP0877090, and Linkage Project Grant LP0991663, and Future Fellowships Grant FT120100487.

## REFERENCES

[1] D. Niyato, E. Hossain, and Z. Han, "Dynamics of multiple-seller and multiple-buyer spectrum trading in cognitive radio networks: A game-theoretic modeling approach," *IEEE Trans. Mobile Comput.*, vol. 8, no. 4, pp. 1009–1022, Aug. 2009.

[2] Z. Ji and R. Liu, "Multi-stage pricing game for collusion-resistant dynamic spectrum allocation," *IEEE J. Select. Areas Commun.*, vol. 26, no. 1, pp. 182–191, Jan. 2008.

[3] S. K. Jayaweera and T. Li, "Dynamic spectrum leasing in cognitive radio networks via primary-secondary user power control games," *IEEE Trans. Wireless Commun.*, vol. 8, no. 6, pp. 3300–3310, Jun. 2009.

[4] P. Lin, J. Jia, Q. Zhang, and M. Hamdi, "Dynamic spectrum sharing with multiple primary and secondary users," *IEEE Trans. Veh. Technol.*, vol. 60, no. 4, pp. 1756–1765, May 2011.

[5] L. Gao, Y. Xu, and X. Wang, "MAP: Multiauctioneer progressive auction for dynamic spectrum access," *IEEE Trans. Mobile Comput.*, vol. 10, no. 8, pp. 1144–1161, Aug. 2011.

[6] V. Gajic, J. Huang, and B. Rimoldi, "Competition of wireless providers for atomic users: Equilibrium and social optimality," in *Proc. 47th Annual Allerton Conference on Communication, Control, and Computing*, 2009.

[7] Y. Tan, S. Sengupta, and K. Subbalakshmi, "Competitive spectrum trading in dynamic spectrum access markets: A price war," in *IEEE Global Telecommunications Conference (GLOBECOM)*, Dec. 2010, pp. 1–5.

[8] H. S. Mohammadian and B. Abolhassani, "Auction-based spectrum sharing for multiple primary and secondary users in cognitive radio networks," in *Proc. IEEE Sarnoff Symposium*, 2010.

[9] D. Li, Y. Xu, J. Liu, X. Wang, and Z. Han, "A market game for dynamic multi-band sharing in cognitive radio networks," in *IEEE International Conference on Communications (ICC)*, vol. 1, May 2010, pp. 1–5.

[10] C. Z. X. Z. M. Z. J. W. Y. Yan, X. Chen, "Bargaining-based spectrum sharing in cognitive radio network."

[11] J. N. Laneman, D. N. C. Tse, and G. W. Wornell, "Cooperative diversity in wireless networks: Efficient protocols and outage behavior," *IEEE Trans. Inform. Theory*, vol. 50, no. 12, pp. 3062–3080, Dec. 2004.

[12] O. Simeone, I. Stanojev, S. Savazzi, Y. Bar-Ness, U. Spagnolini, and R. Pickholtz, "Spectrum leasing to cooperating secondary ad hoc networks," *IEEE J. Select. Areas Commun.*, vol. 26, no. 1, pp. 203–213, Jan. 2008.

[13] L. Giupponi and C. Ibars, "Distributed cooperation among cognitive radios with complete and incomplete information," *EURASIP J. on Advances in Signal Processing*, 2009.

[14] Y. Han, A. Pandharipande, and S. H. Ting, "Cooperative decode-and-forward relaying for secondary spectrum access," *IEEE Trans. Wireless Commun.*, vol. 8, no. 10, pp. 4945–4950, Oct. 2009.

[15] C. H. S. He, L. Jiang, "A novel secondary user assisted relay mechanism in cognitive radio networks with multiple primary users," in *Proc. IEEE Global Telecommunications Conference (GLOBECOM)*, Dec. 2012, pp. 1273–1277.

[16] H. Z. Z. Zhang Z, "A variable-population evolutionary game model for resource allocation in cooperative cognitive relay networks," *IEEE Commun. Lett.*, vol. Early Access Article, Nov. 2013.

[17] Q. Z. J. Zhang, "Stackelberg game for utility-based cooperative cognitive radio networks," in *Proc. of the 10th ACM International Symposium on Mobile Ad Hoc Networking and Computing*, 2009, pp. 23–32.

[18] H. Xu and B. Li, "Efficient resource allocation with flexible channel cooperation in ofdma cognitive radio networks," in *IEEE INFOCOM, 2010*, Mar. 2010, pp. 1–10.

[19] R. Manna, R. H. Y. Louie, Y. Li, and B. Vucetic, "Interference cancellation in two hop multiuser cognitive radio networks," in *Proc. IEEE Wireless Communications and Networking Conference (WCNC)*, Sydney, Australia, Apr. 2010.

[20] R. Manna, , R. H. Y. Louie, Y. Li, and B. Vucetic, "Cooperative spectrum sharing in cognitive radio networks with multiple antennas," *IEEE Trans. Signal Processing*, vol. 59, no. 11, pp. 5509 –5522, Nov. 2011.

[21] J. H. L. Duan, L. Gao, "Cooperative spectrum sharing: A contract-based approach," *IEEE Trans. Mobile Comput.*, vol. Early Access Article.

[22] B. Wang, Y. Wu, Z. Ji, K. Liu, and T. Clancy, "Game theoretical mechanism design methods," *IEEE Trans. Signal Processing*, vol. 25, no. 6, pp. 74–84, Nov. 2008.

[23] J. Zhu and K. Liu, "Belief-assisted pricing for dynamic spectrum allocation in wireless networks with selfish users," in *3rd Annual IEEE Communications Society on Sensor and Ad Hoc Communications and Networks (SECON)*, vol. 1, Sep. 2006, pp. 119 –127.

[24] A. Roth and M. Sotomayor, *Two Sided Matching: a study in game-theoretic modeling and analysis*, 1st ed. UK: Cabridge University Press., 1989.

[25] V. Krishna, *Auction Theory*, 2nd ed. England: Elsevier Inc., 2010.

[26] E. Jorswieck, "Stable matchings for resource allocation in wireless networks," in *17th International Conference on Digital Signal Processing (DSP)*, Jul. 2011, pp. 1 –8.

[27] Y. Leshem and E. Zehavi, "Stable matching for channel access control in cognitive radio systems," in *2nd International Workshop on Cognitive Information Processing (CIP)*, Jun. 2010, pp. 470 –475.

[28] G. Demange, D. Gale, and M. Sotomayor, "Multi-item auctions," *J. Polit. Economy*, no. 94, pp. 863–872, 1986.

[29] S. Bayat, R. H. Y. Louie, Y. Li, and B. Vucetic, "Cognitive radio relay networks with multiple primary and secondary users: Distributed stable matching algorithms for spectrum access," in *Proc. IEEE International Conference on Communications (ICC)*, Kyoto, Japan, Jun. 2011, pp. 1–6.

[30] E. Hossain, D. Niyato, and Z. Han, *Dynamic Spectrum Access and Management in Cognitive Radio Networks*, 1st ed. UK: Cabridge University Press., 2009.

[31] P. Liu, R. Berry, and M. Honig, "A fluid analysis of a utility-based wireless scheduling policy," *IEEE Trans. Inform. Theory*, vol. 52, no. 7, pp. 2872 –2889, Jul 2006.

[32] Z. Jiang, Y. Ge, and Y. Li, "Max-utility wireless resource management for best-effort traffic," *IEEE Trans. Wireless Commun.*, vol. 4, no. 1, pp. 100 – 111, Jan. 2005.

[33] J. W. Lee, M. Chiang, and A. Calderbank, "Price-based distributed algorithms for rate-reliability tradeoff in network utility maximization," *IEEE J. Select. Areas Commun.*, vol. 24, no. 5, pp. 962 – 976, May 2006.

[34] D. Niyato and E. Hossain, "Competitive pricing for spectrum sharing in cognitive radio networks: Dynamic game, inefficiency of nash equilibrium, and collusion," *IEEE J. Select. Areas Commun.*, vol. 26, no. 1, pp. 192–202, Jan. 2008.

[35] P. Y. Chen, W. C. Ao, S. C. Lin, and K. C. Chen, "Reciprocal spectrum sharing game and mechanism in cellular systems with cognitive radio users," in *Proc. IEEE Global Telecommunications Conference (GLOBECOM) Workshops*, Dec. 2011, pp. 981 –985.

[36] X. Kang, R. Zhang, and M. Motani, "Price-based resource allocation for spectrum-sharing femtocell networks: A stackelberg game approach," *IEEE J. Select. Areas Commun.*, vol. 30, no. 3, pp. 538 –549, Apr 2012.

[37] Y. Choi, P. Voltz, and F. Cassara, "On channel estimation and detection for multicarrier signals in fast and selective rayleigh fading channels," *IEEE Trans. Commun.*, vol. 49, no. 8, pp. 1375–1387, Aug 2001.

[38] Q. Zhao and B. Sadler, “A survey of dynamic spectrum access,” *IEEE Trans. Inform. Theory*, vol. 24, no. 3, pp. 79 –89, May 2007.

[39] H. Xu, J. Jin, and B. Li, “A channel portfolio optimization framework for trading in a spectrum secondary market,” in *Proc. IEEE International Conference on Communications (ICC)*, May 2010, pp. 1 –5.

[40] Z. Kong and Y. K. Kwok, “Relay auction algorithms for downlink bandwidth allocation in IEEE 802.16-based OFDM/TDMA wireless mesh networks,” in *Proc. Fourth International Conference on Communications and Networking in China (ChinaCOM)*, Aug. 2009, pp. 1 –5.

[41] D. Gusfield and R. W. Irving, *The Stable Marriage Problem: Structure and Algorithms*. The MIT Press, 1989.

[42] Cisco. Cisco Aironet 1240AG Series 802.11A/B/G Access Point, 802.11a/b/g datasheet, 2009.

[43] A. Kan and J. Telgen, “The complexity of linear programming,” *J. of Statistica Neerlandica*, no. 2, pp. 91–107, 1981.